\documentclass[a4paper,11pt]{article}

\usepackage{natbib}
\usepackage{eqnarray,amsmath,amsthm}
\usepackage{amssymb}
\usepackage{graphicx}
\usepackage{epstopdf}
\usepackage{authblk} 
\usepackage{mathtools}
\usepackage{mathabx}
\usepackage{xcolor,colortbl}
\usepackage{algorithm}
\usepackage{algorithmicx}
\usepackage{algpseudocode}
\usepackage{subfigure}
\usepackage[margin=3cm]{geometry}
\usepackage{nicefrac}
\usepackage{changepage}

\newcommand{\keywords}[1]{\textbf{\textit{Keywords---}} #1}
\newcommand{\norm}[1]{\left\lVert#1\right\rVert}
\newcommand{\vect}[1]{\boldsymbol{#1}}
\newcommand{\params}{\boldsymbol{\theta}}
\newcommand{\vpar}{\boldsymbol{\vartheta}}
\newcommand{\vx}{\boldsymbol{x}}
\newcommand{\vy}{\boldsymbol{y}}
\newcommand{\ceil}[1]{\left\lceil #1\right\rceil}
\newcommand{\easmcda}{DA-SMC$^2$}
\newcommand{\easmcdt}{DT-SMC$^2$}
\newcommand{\esjd}{\textrm{ESJD}}

\makeatletter
\algnewcommand{\LineComment}[1]{{\Statex \hskip\ALG@thistlm \footnotesize\textcolor{blue}{/* #1 */}}}
\algnewcommand{\LineCommentIndent}[1]{{\Statex \hskip\ALG@tlm \footnotesize\textcolor{blue}{/* #1 */}}}
\makeatother

\makeatletter
\newcommand{\algorithmfootnote}[2][\footnotesize]{%
	\let\old@algocf@finish\@algocf@finish
	\def\@algocf@finish{\old@algocf@finish
		\leavevmode\rlap{\begin{minipage}{\linewidth}
				#1#2
		\end{minipage}}%
	}%
}
\makeatother

\makeatletter
\renewcommand\paragraph{\@startsection{paragraph}{4}{\z@}%
	{-2.5ex\@plus -1ex \@minus -.25ex}%
	{1.25ex \@plus .25ex}%
	{\normalfont\normalsize\bfseries}}
\makeatother

\begin{document}
	\setlength{\parindent}{0pc}
	\setlength{\parskip}{1ex}
	
	\title{\bf Automatically adapting the number of state particles in SMC$^2$}

	\author[$1$,$3$,$4$]{Imke Botha} 
	\author[$2$,$3$,$5$]{Robert Kohn}
	\author[$1$,$3$,$4$]{Leah South}
	\author[$1$,$3$,$4$]{Christopher Drovandi}
	
	\affil[$1$]{School of Mathematical Sciences, Queensland University of Technology (QUT)}
	\affil[$2$]{School of Economics, University of New South Wales}
	\affil[$3$]{Australian Research Council Centre of Excellence for Mathematical \& Statistical	Frontiers (ACEMS)}
	\affil[$4$]{QUT Centre for Data Science}
	\affil[$5$]{DARE: ARC training centre in data analytics for reources and environments}
	
	\maketitle
	\begin{abstract}
		Sequential Monte Carlo squared (SMC$^2$) methods can be used for parameter inference of intractable likelihood state-space models. These methods replace the likelihood with an unbiased particle filter estimator, similarly to particle Markov chain Monte Carlo (MCMC). As with particle MCMC, the efficiency of SMC$^2$ greatly depends on the variance of the likelihood estimator, and therefore on the number of state particles used within the particle filter. We introduce novel methods to adaptively select the number of state particles within SMC$^2$ using the expected squared jumping distance to trigger the adaptation, and modifying the exchange importance sampling method of \citet{Chopin2012a} to replace the current set of state particles with the new set of state particles. The resulting algorithm is fully automatic, and can significantly improve current methods. Code for our methods is available at https://github.com/imkebotha/adaptive-exact-approximate-smc.
	\end{abstract}

	
	\maketitle

	\section{Introduction} \label{sec:intro}
	
	We are interested in exact Bayesian parameter inference for state-space models (SSMs) where the likelihood function of the model parameters is intractable. SSMs are ubiquitous in engineering, econometrics and the natural sciences; see \citet{Cappe2005} and references therein for an overview. They are used when the process of interest is observed indirectly over time or space, i.e.\ they consist of a hidden or latent process $\{X_t\}_{t\ge 1}$ and an observed process $\{Y_t\}_{t\ge 1}$. 
	
	Particle Markov chain Monte Carlo \citep[MCMC;][]{Andrieu2010a,Andrieu2009a} methods such as particle marginal Metropolis-Hastings (PMMH) or particle Gibbs can be used for exact parameter inference of intractable likelihood SSMs. PMMH uses a particle filter estimator of the likelihood within an otherwise standard Metropolis-Hastings algorithm. Similarly, particle Gibbs uses a conditional particle filter to draw the latent states from their full conditional distribution, then updates the model parameters conditional on the latent states. Both PMMH and particle Gibbs are simulation consistent under mild conditions \citep{Andrieu2010a}.

	\citet{Chopin2012a} and \citet{Duan2014a} apply a similar approach to sequential Monte Carlo (SMC) samplers. SMC methods for static models \citep{Chopin2002, DelMoral2006} recursively sample through a sequence of distributions using a combination of reweighting, resampling and mutation steps. In the Bayesian setting, this sequence often starts at the prior and ends at the posterior distribution. For intractable likelihood SSMs, \citet{Chopin2012a} and \citet{Duan2014a} replace the likelihood within the sequence of distributions being traversed with its unbiased estimator. Practically, this means that each parameter particle is augmented with $N_x$ state particles. Due to this nesting of SMC algorithms and following \citet{Chopin2012a}, we refer to these methods as SMC$^2$. As with particle MCMC, for any fixed number of state particles ($N_x$), SMC$^2$ targets the exact posterior distribution \citep{Duan2014a}. 
	
	While other, similar methods are available for Bayesian parameter inference of intractable likelihood SSMs, e.g.\ nested particle filters \citep{Crisan2017,Crisan2018} and ensemble MCMC \citep{Drovandi2022}, the resulting inference is approximate and so is not considered in this paper. 
	
	The sampling efficiency of particle MCMC and SMC$^2$ greatly depends on the number of state particles used within the particle filter. In particle MCMC, $N_x$ is generally tuned manually, which can be time intensive. A significant advantage of SMC$^2$ over particle MCMC is that $N_x$ can be adapted automatically. Strategies to do this are proposed by \citet{Chopin2012a,Chopin2015} and \citet{Duan2014a}; however, these methods automate the adaptation of $N_x$ at the expense of other model-specific tuning parameters, which must then be tuned manually. Furthermore, the value of $N_x$ can be difficult to choose in practice, and has a significant effect on both the Monte Carlo error of the SMC approximation to the target distribution and the computation time. Current methods require a moderate starting value of $N_x$ to avoid poor values in subsequent iterations, i.e.\ values that are too low and negatively impact the accuracy of the samples, or unnecessarily high values that increase the computation time. 
	
	Our article introduces a novel and principled strategy to automatically tune $N_x$, while aiming to keep an optimal balance between statistical and computational efficiency. Compared to current methods, our approach has less tuning parameters that require manual calibration. We find that using the expected squared jumping distance of the mutation step to adapt the number of state particles generally gives the most efficient and reliable results. To further improve the overall efficiency of the adaptation, we also modify the exchange importance sampling method of \citet{Chopin2012a} to update the set of state particles once $N_x$ is adapted. This modified version introduces no extra variability in the parameter particle weights, and outperforms the current methods.

	The rest of the paper is organized as follows. Section \ref{sec:smc} gives the necessary background on state-space models and SMC methods, including particle filters, SMC for static models and SMC$^2$. Section \ref{sec:easmc_adapting} describes the current methods for adapting the number of state particles in SMC$^2$. Section \ref{sec:methods} describes our novel tuning methodology. Section \ref{sec:ex} shows the performance of our methods on a Brownian motion model, a stochastic volatility model,  a noisy theta-logistic model and a noisy Ricker model. Section \ref{sec:disc} concludes.
	
	\section{Background} \label{sec:smc}
	This section contains the necessary background information for understanding the novel methods discussed in Section \ref{sec:methods}. It covers content related to exact Bayesian inference for state-space models, particularly focussed on models with intractable transition densities.
	
	\subsection{State-Space Models}
	Consider a state-space model (SSM) with parameters $\params \in \Theta$, a hidden or latent process $\{X_t\}_{t\ge 1}$ and an observed process $\{Y_t\}_{t\ge 1}$. A key assumption of SSMs is that the process $\{(X_t, Y_t), t\ge 1\}$ is Markov, and we further assume that the full conditional densities of $Y_t = y_t$ and $X_t = x_t$ are
	\begin{align*}
		p(y_t\mid x_t, x_{t-1}, y_{t-1},\params) = g(y_t\mid x_t, \params),
	\end{align*}
	and 
	\begin{align*}
		p(x_t \mid x_{t-1}, y_{t-1}, \params) = f(x_t \mid x_{t-1}, \params),
	\end{align*}
	where $g(y_t\mid x_t, \vect\theta)$ and $f(x_t\mid x_{t-1}, \vect\theta)$ are the observation density and transition density respectively. The density of the latent states at time $t=1$ is $\mu(x_1\mid\params)$ and the prior density of the parameters is $p(\params)$. 

	Define $\vect{z}_{i:j} \coloneqq \{z_i, z_{i+1}, \ldots, z_j\}$ for $j\ge i$. The distribution of $\params$ conditional on the observations up to time $t \le T$ is
	\begin{align}
		p(\params\mid\vy_{1:t})	=\frac{p(\params)}{p(\vy_{1:t})} \int_{\vx_{1:t}}{
			p(\vx_{1:t}, \vy_{1:t}\mid \params)
		}d\vx_{1:t}, 
		\label{eqn:marginal_posterior}
	\end{align}
	where 
	\begin{align}
		p(\vx_{1:t}, \vy_{1:t}\mid \params) = \mu(x_1\mid\params)\prod_{i=2}^t{f(x_i \mid x_{i-1}, \params)}\prod_{i=1}^t{g(y_i\mid x_i, \params)}.
		\label{eqn:posterior}
	\end{align}
	The integral in \eqref{eqn:marginal_posterior} gives the likelihood function $p(\vy_{1:t}\mid\params)$. This integral is often analytically intractable or prohibitively expensive to compute, which means that the likelihood is also intractable. If the value of $\params$ is fixed, a particle filter targeting $p(\vx_{1:t}\mid \vy_{1:t}, \params)$ gives an unbiased estimate of the likelihood as a by-product, as described in Section \ref{sec:pf}. Similarly, a conditional particle filter \citep{Andrieu2010a}, i.e.\ a particle filter that is conditional on a single state trajectory $\vx_{1:t}^k$, can be used to unbiasedly simulate latent state trajectories from $p(\cdot\mid \vx_{1:t}^k, \vy_{1:t}, \params)$. Particle filters are SMC methods applied to dynamic models.
	
	\subsection{Sequential Monte Carlo}
	SMC methods recursively sample from a sequence of distributions,
	$\pi_d(z_d) \propto \gamma_d(z_d)$, $d = 0, \ldots, D$, where $\pi_0(z_0)$ can generally be sampled from directly and $\pi_{D}(z_D)$ is the target distribution \citep{DelMoral2006}. 
	
	These distributions are traversed using a combination of resample, mutation and reweight steps. Initially, $N_{z}$ samples are drawn from $\pi_0(z_0)$ and given equal weights $\{z_0^{n}, W_0^n=\nicefrac{1}{N_{z}}\}_{n=1}^{N_{z}}$. For each subsequent distribution, the particles are resampled according to their weights, thus removing particles with negligible weights and duplicating high-weight particles. The resampled particles are then mutated using $R$ applications of the mutation kernel $K(z^n_{d-1},z^n_{d})$, and reweighted as
	\begin{align*}
		w_d^n =  N_{z}^{-1}\cdot\frac{\gamma_{d}(z^n_{d})L(z^n_{d}, z^n_{d-1})}{\gamma_{d-1}(z^n_{d-1})K(z^n_{d-1}, z^n_{d})}, \quad W_d^n = \frac{w_d^n}{\sum_{i=1}^{N_{z}}w_d^i},
	\end{align*}
	where $L(z^n_{d}, z^n_{d-1})$ is the artificial backward kernel of \citet{DelMoral2006}. Note that if the weights at iteration $d$ are independent of the mutated particles $z^n_{d}$, the reweighting step should be completed prior to the resample and mutation steps. At each iteration $d$, the weighted particles $\{z_d^n, W_d^n\}_{n=1}^{N_{z}}$ form an approximation of $\pi_d(z_d)$. See \citet{DelMoral2006} for more details.
	
	An advantage of SMC methods is that an unbiased estimate of the normalizing constant of the target distribution can be obtained as follows \citep{DelMoral2006}
	\begin{align}
		\int{\gamma_{D}(z_D)} dz \approx \prod_{d=0}^{D}{\sum_{n=1}^{N_{z}}{w_d^{n}}}.
		\label{eqn:llest}
	\end{align}
	This feature is exploited in the SMC$^2$ methods described in Section \ref{sec:smc2}.
	
	\subsubsection{Particle Filters} \label{sec:pf}
	SMC methods for dynamic models are known as particle filters. For fixed $\params$, the sequence of filtering distributions for $d = 1, \ldots, T$ is
	\begin{align*}
		\pi_d(z_d) := p(\vx_{1:d} \mid \vy_{1:d}, \params) = \frac{\mu(x_1\mid\params)}{p(\vy_{1:d}\mid\params)}\prod_{i=2}^d{f(x_i \mid x_{i-1}, \params)}\prod_{i=1}^d{g(y_i\mid x_i, \params)}.
	\end{align*}
	The bootstrap particle filter of \citet{Gordon1993} uses the transition density as the mutation kernel $K(x_{d-1},x_{d}) = f(x_d \mid x_{d-1}, \params)$, and selects $L(x_{d},x_{d-1}) = 1$ as the backward kernel. The weights are then given by
	\begin{align*}
		w_d^m = N_{x}^{-1}g(y_d\mid x_d, \params), \quad W_d^m = \frac{w_d^m}{\sum_{i=1}^{N_{x}}w_d^i},
	\end{align*}
	for $m = 1, \ldots, N_x$. Algorithm \ref{alg:BPF} shows pseudo-code for the bootstrap particle filter \citep{Gordon1993}. 
	
	Define $x_{1:d}^{1:N_x}:=\{x_1^{1:N_x}, \dots, x_d^{1:N_x}\}$, where $d=1, \ldots, T$. The likelihood estimate with $N_x$ state particles and $d$ observations is then 
	\begin{align}
		\begin{split}
			\widehat{p_{N_x}}(\vy_{1:d}\mid\params, \vx_{1:d}^{1:N_x}) = \prod_{i=1}^{d}{\sum_{m=1}^{N_{x}}{w_i^{m}}} 
			= \prod_{i=1}^{d}{\left(
				\frac{1}{N_x} \sum_{m=1}^{N_x}g(y_i\mid x_i^m, \params)
				\right)}.
		\end{split}
		\label{eq:lest}
	\end{align}
	Let $\psi(\vx_{1:d}^{1:N_x})$ be the joint distribution of all the random variables drawn during the course of the particle filter \citep{Andrieu2010a}. The likelihood estimate in \eqref{eq:lest} is unbiased in the sense that $\mathbb{E}_{\psi(\vx_{1:d}^{1:N_x})}\left(\widehat{p_{N_x}}(\vy_{1:d}\mid\params, \vx_{1:d}^{1:N_x})\right) = p(\vy_{1:d}\mid\params)$ \citep[Section 7.4.2 of \citealp{DelMoral2004}; see also][]{Pitt2012}. 
	
	The notation 
	\begin{align*}
		\widehat{p_{N_x}}(\vy_{1:d}\mid\params) &=  
		\widehat{p_{N_x}}(\vy_{1:d},\vx_{1:d}^{1:N_x}\mid\params) = \widehat{p_{N_x}}(\vy_{1:d}\mid\params, \vx_{1:d}^{1:N_x})\psi(\vx_{1:d}^{1:N_x}) \\
		&= \frac{1}{N_x} \sum_{m=1}^{N_x}{\widehat{p_{N_x}}(\vy_{1:d}\mid\params, \vx_{1:d}^{m})}, \quad \vx_{1:d}^{m} \sim \psi(\vx_{1:d}^{m}),
	\end{align*}
	is used interchangeably throughout the paper.
	
	\begin{algorithm}[htp]
		\small
		\begin{adjustwidth}{\algorithmicindent}{}
			\textbf{Input: } data $\vect{y}_{1:d}$, number of state particles $N_x$ and the static parameters $\params$. \\
			\textbf{Output: } likelihood estimate $\widehat{p_{N_x}}(\vy_{1:d}\mid\params)$, set of weighted state particles $\{\vx_{1:d}^{1:N_x}, \vect{W}_{1:d}^{1:N_x}\}$
		\end{adjustwidth}
		\vspace{0.5em}
		
		\begin{algorithmic}[1]
			
			\LineComment{Initialise (t=1)}
			\State Initialise $x_1^{1:N_x}\sim\mu(\cdot\mid\params)$ and calculate the initial weights 
			\begin{align*}
				w_1^{(m)} = N_x^{-1}\cdot g(y_1 \mid x_{1}^{(m)}, \params), \quad W_1^{(m)}=\frac{w_1^{(m)}}{\sum_{i=1}^{N_x}{w_1^{i}}}
			\end{align*}
			
			\LineComment{Initialise likelihood estimate}
			\State Initialise the likelihood estimate $\widehat{p_{N_x}}(\vy_{1}\mid\params)=\sum_{m=1}^{N_x}{w_1^{m}}$
			\vspace{0.5em}
			
			\For{$t=2$ to $d$}
			\LineCommentIndent{Resample}
			\State Resample $N_x$ particles from $\vx_{t-1}^{1:N_x}$ with probability $\vect{W}_{t-1}^{1:N_x}$
			\vspace{0.5em}
			
			\LineComment{Simulate forward}
			\State Simulate the particles forward, $x_t^{(m)}\sim f(\cdot\mid x_{t-1}^{(m)}, \params)$ 
			\vspace{0.5em}
			
			\LineComment{Reweight}
			\State Re-weight the particles from $\pi_{t-1}(\cdot)$ to $\pi_{t}(\cdot)$ 
			\begin{align*}
				w_t^{(m)} = \frac{1}{N_x}\cdot g(y_t \mid x_{t}^{(m)}, \params), \quad W_t^{(m)}=\frac{w_t^{(m)}}{\sum_{i=1}^{N_x}{w_t^{i}}}
			\end{align*}
			
			\LineComment{Update likelihood estimate}
			\State Update the likelihood estimate $\widehat{p_{N_x}}(\vy_{1:t}\mid\params) = \widehat{p_{N_x}}(\vy_{1:t-1}\mid\params) \cdot \sum_{m=1}^{N_x}{w_t^{m}}$ 
			\vspace{0.5em}
			\EndFor
			
		\end{algorithmic}
		\caption{The bootstrap particle filter of \citet{Gordon1993}. The index $(m)$ means `for all $m\in \{1,\ldots,N_x\}$'}
		\label{alg:BPF}
	\end{algorithm}
	
	\subsubsection{SMC for Static Models}
	For static models, where inference on $\params$ is of interest, the sequence of distributions traversed by the SMC algorithm is $\pi_d(\params_d) \propto \gamma_d(\params_d)$, $d = 0, \ldots, D$, where $\pi_0(\params_0) = p(\params)$ is the prior and $\pi_{D}(\params_D) = p(\params\mid\vy_{1:T})$ is the posterior distribution. Assuming that the likelihood function is tractable, there are at least two general ways to construct this sequence,
	\begin{enumerate}
		\item likelihood tempering, which gives $\pi_d(\params) \ \propto \ p(\vy_{1:T}\mid\vect{\theta})^{g_d}p(\params)$ for $d=0,\ldots,D$, and where $0 = g_0 \le \cdots \le g_{D} = 1$, and
		\item data annealing \citep{Chopin2002}, which gives $\pi_d(\vect\theta) \ \propto \ p(\vy_{1:d}\mid\vect{\theta})p(\vect{\theta})$ for $d = 0,\ldots,T$, where $T$ is the number of observations and $D=T$.
	\end{enumerate}
	Typically, SMC for static models uses a mutation kernel which ensures that the current target $\pi_d(\vect\theta)$ remains invariant. A common choice is to use $R$ applications of an MCMC mutation kernel along with the backward kernel $L(\params_{d}, \params_{d-1}) = \gamma_d(\params_{d-1})K(\params_{d-1}, \params_{d})\slash \gamma_d(\params_{d})$ \citep{Chopin2002, DelMoral2006}. The weights then become
	\begin{align}
		w_d^n = N_{\theta}^{-1}\cdot\frac{\gamma_d(\params_{d-1}^{n})}{\gamma_{d-1}(\params^n_{d-1})}, \quad	W_d^n = \frac{w_d^n}{\sum_{i=1}^{N_{\theta}}w_d^i}. \label{eqn:smc_weights}
	\end{align}
	Since the weights are independent of the mutated particles $\params_{d}$, the reweighting step is completed prior to the resample and mutation steps.
	
	\subsection{SMC$^2$} \label{sec:smc2}
	Standard SMC methods for static models cannot be applied directly to state-space models if the parameters $\params$ are unknown except when the integral in \eqref{eqn:marginal_posterior} is analytically tractable. When the likelihood is intractable, SMC$^2$ replaces it in the sequence of distributions being traversed with a particle filter estimator. Essentially, each parameter particle is augmented with a set of weighted state particles. 
	
	Since the likelihood is replaced with a particle filter estimator, the parameter particles in SMC$^2$ are mutated using $R$ applications of a particle MCMC mutation kernel $K(\cdot, \cdot)$. Section \ref{sec:pmmh} describes the particle marginal Metropolis-Hastings (PMMH) algorithm. As with SMC for static models, the parameter particle weights are given by \eqref{eqn:smc_weights}. 

	Two general ways to construct the sequence of targets for SMC$^2$ are the density tempered marginalised SMC algorithm of \citet{Duan2014a} and the data annealing SMC$^2$ method of \citet{Chopin2012a}, which we refer to as density tempering SMC$^2$ (\easmcdt{}) and data annealing SMC$^2$ (\easmcda{}) respectively. These are described in Sections \ref{sec:dtsmc2} and \ref{sec:dasmc2}. 
	
	Algorithm \ref{alg:easmc} shows pseudo-code which applies to both \easmcdt{} and \easmcda{}. The main difference between the two methods is how the sequence of targets is defined. Sections \ref{sec:dtsmc2} and \ref{sec:dasmc2} describe the sequence of targets and the reweighting formulas for \easmcdt{} and \easmcda{} respectively. For conciseness, we denote the set of weighted state particles associated with parameter particle $n$, $n = 1,\ldots, N_{\theta}$ at iteration $d$ as 	
	\begin{align*}
		\tilde{\vx}_d^{1:N_x, n} := 
		\begin{cases}
			\{\vx_{1:d}^{1:N_x,n}, \vect{S}_{d}^{1:N_x,n}\}, & \text{for \easmcda{}}, \\
			\{\vx_{1:T}^{1:N_x,n}, \vect{S}_{d}^{1:N_x,n}\}, & \text{for \easmcdt{}},
		\end{cases}
	\end{align*}
	where $\vect{S}_{d}^{1:N_x,n}$ is the set of normalised state particle weights. The $n$th parameter particle with its attached set of weighted state particles is denoted as $\vect{\vartheta}_d^n = \{\params_d^n, \tilde{\vx}_d^{1:N_x, n}\}$, $n = 1,\ldots, N_{\theta}$. 
	
	\subsubsection{Density Tempering SMC$^2$} \label{sec:dtsmc2}
	The sequence of distributions for \easmcdt{} is 
	\begin{align*}
		\pi_d(\params) \propto p(\params)\left[\widehat{p_{N_x}}(\vy_{1:T}\mid\params, \vx_{1:T}^{1:N_x})\right]^{g_d}\psi(\vx_{1:T}^{1:N_x}), \quad 0=g_0\le\cdots\le g_D=1,
	\end{align*}
	which gives the weights from \eqref{eqn:smc_weights} as
	\begin{align}
		w_d^{n} = N_{\theta}^{-1} \cdot \left[\widehat{p_{N_x}}(\vy_{1:T}\mid\params_{d-1}^n, \vx_{1:T}^{1:N_x})\right]^{g_{d}- g_{d-1}}, \quad W_d^n = \frac{w_d^{n}}{\sum_{i=1}^{N_{\theta}}{w_d^{i}}}. \label{eqn:dtsmc2_weights}
	\end{align}
	Due to the tempering parameter $g_d$, \easmcdt{} is only exact at the first and final temperatures, i.e.\ $p(\params)p(\vy_{1:T}\mid\params)^{g_d}\slash \int{p(\params)p(\vy_{1:T}\mid\params)^{g_d}}d\params$ is a marginal distribution of $\pi_d(\params)$ only at $g_1 = 0$ and $g_D = 1$. 
	
	\subsubsection{Data Annealing SMC$^2$} \label{sec:dasmc2}
	For \easmcda{}, the sequence of distributions is 
	\begin{align*}
		\pi_d(\params) \
		&\propto \ p(\params) \widehat{p_{N_x}}(\vy_{1:d}\mid\params, \vx_{1:d}^{1:N_x})\psi(\vx_{1:d}^{1:N_x}), \quad D=T,
	\end{align*}
	and the weights from \eqref{eqn:smc_weights} are
	\begin{align}
		w_d^{n} = N_{\theta}^{-1} \cdot \widehat{p_{N_x}}\left(y_{d}\mid \vect{y}_{1:d-1}, \vect{\theta}_{d-1}^{n}\right),\quad W_d^n = \frac{w_d^{n}}{\sum_{i=1}^{N_{\theta}}{w_d^{i}}}, \label{eqn:dasmc2_weights}
	\end{align}
	where $\widehat{p_{N_x}}\left(y_{d}\mid \vect{y}_{1:d-1}, \vect{\theta}_{d-1}^{n}\right)$ is obtained from iteration $d$ of a particle filter (see \eqref{eq:lest} and Algorithm \ref{alg:BPF}). Unlike \easmcdt{}, \easmcda{} admits $p\left(\params\mid\vy_{1:d}\right)$ as a marginal distribution of $\pi_d(\params)$ for all $d=0,\ldots,D$. 
	
	\begin{algorithm}[htp]
		\small
		\begin{adjustwidth}{\algorithmicindent}{}
			\textbf{Input: } data $\vy_{1:T}$, number of parameter particles $N_{\theta}$, number of state particles $N_x$, number of MCMC iterations $R$  \\
			\textbf{Output: } set of weighted particles $\{\vpar_D^{1:N_{\params}}, \vect{W}_D^{1:N_{\params}}\}$
		\end{adjustwidth}
		\vspace{0.5em}
		
		\begin{algorithmic}[1]
			\LineComment{Initialisation step (t=0)}
			
			\State Initialise $\vpar_0^{1:N_{\theta}}$ and set $W_0^{(n)} = \frac{1}{N_{\theta}}$
			\vspace{0.5em}
			
			\For{$d=1$ to $D$} 
			\LineCommentIndent{Reweight}
			\State Re-weight the particles from $\pi_{d-1}(\cdot)$ to $\pi_{d}(\cdot)$ using \eqref{eqn:dtsmc2_weights} or \eqref{eqn:dasmc2_weights}.
			\vspace{0.5em}
			
			\LineComment{Resample}
			\State Resample $N_{\vect{\theta}}$ particles from $\vpar_d^{1:N_{\theta}}$ with probability $\vect{W}_{d}^{1:N_{\theta}}$ 
			
			\vspace{0.5em}
			\LineComment{Mutate}
			
			\For{$r=1$ to $R$} 
			\State PMMH mutation $\vpar_{d}^{(n)}\sim K\left(\vpar_{d}^{(n)}, \cdot\right)$ (See Algorithm \ref{alg:pmmh})
			\EndFor
			\EndFor
		\end{algorithmic}
		\caption{The SMC$^2$ Algorithm. The index $(n)$ means `for all $n\in \{1,\ldots,N_{\theta}\}$'}
		\label{alg:easmc}
	\end{algorithm}

	\subsection{Particle MCMC mutations} \label{sec:pmmh}
	
	The simplest mutation of the parameter particles in SMC$^2$ is a sequence of Markov move steps using the PMMH algorithm; see \citet{Gunawan2021} for alternatives. The PMMH method is a standard Metropolis-Hastings algorithm where the intractable likelihood is replaced by the particle filter estimate in \eqref{eq:lest}. Algorithm \ref{alg:pmmh} shows a single PMMH iteration. 
	
	\begin{algorithm}[htp]
		\small
		\begin{adjustwidth}{\algorithmicindent}{}
			\textbf{Input: } data $\vy$, proposal distribution $q(\cdot)$, current parameter value $\params_d$, current likelihood estimate $\widehat{p_{N_x}}(\vy\mid \params_d)$. Note that $\vy := \vy_{1:T}$ for \easmcdt and $\vy := \vy_{1:d}$ for \easmcda. \textit{Optional:} current set of weighted state particles $\tilde{\vx}_d^{1:N_x}$ \\
			\textbf{Output: } new parameter value $\params_d$, new likelihood estimate $\widehat{p_{N_x}}(\vy\mid \params_d)$. \textit{Optional:} new set of weighted state particles $\tilde{\vx}_d^{1:N_x}$ 
		\end{adjustwidth}
		\vspace{0.5em}
		
		\begin{algorithmic}[1]
			\State Sample $\params_d^*\sim q(\cdot\mid\params_d)$, 
			\vspace{0.5em}
			\State Run Algorithm \ref{alg:BPF} to obtain  $\widehat{p_{N_x}}(\vy\mid\params_d^*)$ and $\tilde{\vx}_d^{1:N_x, *}$, 
			\vspace{0.5em}
			
			\State Calculate acceptance probability 
			\begin{align}
				\alpha(\params_d, \params_d^*) = \min\left(1,\ \frac{\widehat{p_{N_x}}(\vy\mid\params_d^*)p(\params_d^*)}{\widehat{p_{N_x}}(\vy\mid\params_d)p(\params_d)}
				\frac{q(\params_d\mid\params_d^*)}{q(\params_d^*\mid\params_d)}\right).
				\label{eqn:pmcmc_alpha}
			\end{align}
			
			\State With probability $\alpha(\params_d, \params_d^*)$, set
			\begin{align*}
				\params_d=\params_d^*, \quad
				\widehat{p_{N_x}}(\vy\mid\params_d)=\widehat{p_{N_x}}(\vy\mid\params_d^*), \quad \tilde{\vx}_d^{1:N_x} = \tilde{\vx}_d^{1:N_x, *},
			\end{align*} 
			otherwise keep the current values of $\params_d$, $\widehat{p_{N_x}}(\vy\mid\params_d)$ and $\tilde{\vx}_d^{1:N_x}$.	
			
		\end{algorithmic}
		\caption{A single iteration of the particle marginal Metropolis-Hastings algorithm.}
		\label{alg:pmmh}
	\end{algorithm}
	
	While a PMMH mutation leaves the current target invariant, its acceptance rate is sensitive to the variance of the likelihood estimator \citep{Andrieu2010a}. In practice, this means that if the variance is too high, then some particles may not be mutated during the mutation step --- even with a large number of MCMC iterations. 
	
	In the context of particle MCMC samplers, \citet{Andrieu2010a} show that $N_x$ must be chosen as $\mathcal{O}(T)$ to achieve reasonable acceptance rates, i.e.\ reasonable variance of the likelihood estimator. \citet{Pitt2012}, \citet{Doucet2015} and \citet{Sherlock2015} recommend choosing $N_x$ such that the variance of the log-likelihood estimator is between $1$ and $3$ when evaluated at, e.g., the posterior mean. This generally requires a (potentially time-consuming) tuning process for $N_x$ before running the algorithm. 
	
	For SMC$^2$, fewer particles may be required to achieve reasonable acceptance rates in the early stages of the algorithm. In \easmcda{}, $N_x = \mathcal{O}(t)$, where $t=d$, suggests starting with a small $N_x$, and increasing it with each added observation. Likewise, in \easmcdt{}, a small $g_d$ will reduce the impact of a highly variable log-likelihood estimator. In addition, unlike particle MCMC methods, it is possible to automatically adapt $N_x$ within SMC$^2$. The next section describes the tuning strategies proposed by \citet{Chopin2012a,Chopin2015} and \citet{Duan2014a}.
	
	\section{Existing methods to calibrate $N_x$} \label{sec:easmc_adapting}
	
	There are three main stages to adapting $N_x$: (1) triggering the adaptation, (2) choosing the new number of particles $N_x^*$, and (3) replacing the current set of state particles $\tilde{\vx}^{1:N_x, 1:N_{\theta}}_d$ with the new set $\tilde{\vx}^{1:N_x^*, 1:N_{\theta}}_d$. To simplify notation, we write $\tilde{\vx}^{1:N_x, 1:N_{\theta}}_d$ as
	$\tilde{\vx}^{1:N_x}_d$. 
	
	\subsection*{Stage 1. Triggering the adaptation} 
	It may be necessary to adapt $N_x$ when the mutation step no longer achieves sufficient particle diversity. \citet{Chopin2012a,Chopin2015} and \citet{Duan2014a} fix the number of MCMC iterations ($R$) and change $N_x$ whenever the acceptance rate of a single MCMC iteration falls below some target value. This approach has two main drawbacks. First, the acceptance rate does not take the jumping distances of the particles into account, and can be made artificially high by making very local proposals. Second, both $R$ and the target acceptance rate must be tuned --- even if the exact likelihood is used, the acceptance rate may naturally be low, depending on the form of the posterior and the proposal function used within the mutation kernel. Ideally, $N_x$ and $R$ should be jointly adapted. 
	
	\subsection*{Stage 2. Choosing the new number of particles $N_x^*$} 
	A new number of state particles ($N_x^*$) is determined in the second stage.  \citet{Chopin2012a} set $N_x^* = 2\cdot N_x$ (\textsc{double}), while \citet{Duan2014a} set $N_x^* = \widehat{\sigma_{N_x}}^2 \cdot N_x$ (\textsc{rescale-var}), where $\widehat{\sigma_{N_x}}^2$ is the estimated variance of the log-likelihood estimator using $N_x$ state particles. The variance is estimated from $k$ independent estimates of the log-likelihood (for the current SMC target) based on the sample mean of the parameter particles. This choice is motivated by the results of \citet{Pitt2012}, \citet{Doucet2015} and \citet{Sherlock2015}, who show that $\sigma_{N_x}^2 \ \propto \ 1\slash N_x$ for any number of state particles $N_x$. Setting $\sigma_{N_x}^2 = \alpha \slash N_x$ and rearranging gives both $\alpha = \sigma_{N_x}^2 \cdot N_x$ and $N_x = \alpha\slash\sigma_{N_x}^2$. Given $N_x$ and $\sigma_{N_x}^2$, these expressions can be used to find a new number of state particles $N_x^*$ such that $\sigma_{N_x^*}^2 = 1$, by noting that $N_x^* = \alpha\slash\sigma_{N_x^*}^2 = \alpha\slash 1 = \sigma_{N_x}^2 \cdot N_x$.	
	
	We find that if the initial $N_x$ is too small, then the \textsc{double} scheme of \citet{Chopin2012a} can take a significant number of iterations to set $N_x$ to a reasonable value. It can also increase $N_x$ to an unnecessarily high value if the adaptation is triggered when the number of state particles is already large. 
	
	While the \textsc{rescale-var} method of \citet{Duan2014a} is more principled, as it takes the variance of the log-likelihood estimator into account, we find that it is also sensitive to the initial number of particles. For a poorly chosen initial $N_x$, the variance of the log-likelihood estimator can be of order $10^2$ or higher. In this case, scaling the current number of particles by $\widehat{\sigma_{N_x}}^2$ may give an extremely high value for $N_x^*$. 
	
	\citet{Chopin2015} propose a third method; they set $N_x^* = \tau/\sigma^2_{N_x}$, where $\tau$ is a model-specific tuning parameter, and $\sigma^2_{N_x}$ is the variance of the log-likelihood estimator with $N_x$ state particles. This choice is motivated by the results from \citet{Doucet2012arxiv} (an earlier version of \citet{Doucet2015}). See \citet{Chopin2015} for further details. Since the parameter $\tau$ must be tuned manually, this approach is not included in our numerical experiments in Section \ref{sec:ex}. 
	
	\subsection*{Stage 3. Replacing the state particle set} 
	The final stage replaces the current set of state particles $\tilde{\vx}^{1:N_x}_d$ by the new set $\tilde{\vx}^{1:N_x^*}_d$. \citet{Chopin2012a} propose a reweighting step for the parameter particles (\textsc{reweight}) using the generalised importance sampling method of \citet{DelMoral2006} to swap $\tilde{\vx}^{1:N_x}_d$ with $\tilde{\vx}^{1:N_x^*}_d$. The incremental weight function for this step (for \easmcda{}) is 
	\begin{align*}
		IW &= \frac{
			\pi_d\left(\params_d, \vx_{d}^{1:N_x^*} \mid\vy_{1:d}\right)L_d(\vx_{d}^{1:N_x^*}, \vx_{d}^{1:N_x})
		}{
			\pi_d\left(\params_d, \vx_{d}^{1:N_x} \mid\vy_{1:d}\right)\psi(\vx_{d}^{1:N_x^*})
		} \\
		&= \frac{
			p(\params_d) \widehat{p_{N_x^*}}(\vy_{1:d}\mid\params_d, \vx_{d}^{1:N_x^*}) \psi(\vx_{d}^{1:N_x^*})
			L_d(\vx_{d}^{1:N_x^*}, \vx_{d}^{1:N_x})
		}{
			p(\params_d) \widehat{p_{N_x}}(\vy_{1:d}\mid\params_d, \vx_{d}^{1:N_x}) \psi(\vx_{d}^{1:N_x})
			\psi(\vx_{d}^{1:N_x^*})
		} \\
		&= \frac{
			\widehat{p_{N_x^*}}(\vy_{1:d}\mid\params_d, \vx_{d}^{1:N_x^*})
			L_d(\vx_{d}^{1:N_x^*}, \vx_{d}^{1:N_x})
		}{
			\widehat{p_{N_x}}(\vy_{1:d}\mid\params_d, \vx_{d}^{1:N_x}) \psi(\vx_{d}^{1:N_x})
		},
	\end{align*}
	where $L_d(\vx_{d}^{1:N_x^*}, \vx_{d}^{1:N_x})$ is the backward kernel. They use the following approximation to the optimal backward kernel (see Proposition 1 of \citet{DelMoral2006})
	\begin{align}
		L_d(\vx_{d}^{1:N_x^*}, \vx_{d}^{1:N_x}) &= \frac{
			\widehat{p_{N_x}}(\vy_{1:d}\mid\params_d, \vx_{d}^{1:N_x}) \psi(\vx_{d}^{1:N_x})
		}{
			p(\vy_{1:d}\mid\params_d)
		} \nonumber \\ &\approx \frac{
			\widehat{p_{N_x}}(\vy_{1:d}\mid\params_d, \vx_{d}^{1:N_x}) \psi(\vx_{d}^{1:N_x})
		}{
			\widehat{p_{N_x}}(\vy_{1:d}\mid\params_d, \vx_{d}^{1:N_x})
		} = \psi(\vx_{d}^{1:N_x}),
		\label{eq:Lopt}
	\end{align}
	leading to
	\begin{align*}
		IW_d = \frac{\widehat{p_{N_x^*}}(\vy_{1:d}\mid\params_d, \vx_{d}^{1:N_x^*})}{\widehat{p_{N_x}}(\vy_{1:d}\mid\params_d, \vx_{d}^{1:N_x})}.
	\end{align*}
	For density tempering, this becomes
	\begin{align*}
		IW_d = \left(\frac{\widehat{p_{N_x^*}}(\vy_{1:T}\mid\params_d, \vx_{d}^{1:N_x^*})}{\widehat{p_{N_x}}(\vy_{1:T}\mid\params_d, \vx_{d}^{1:N_x})}\right)^{g_d}.
	\end{align*}
	The new parameter particle weights are then given by 
	\begin{align*}
		w_d^n = W_{d-1}^n\cdot IW_{d}^{n}, \quad W_d^n = \frac{w_d^n}{\sum_{i=1}^{N_{\theta}}{w_d^i}}.
	\end{align*}
	While this method is relatively fast, it can significantly increase the variance of the parameter particle weights \citep{Duan2014a}.
	
	As an alternative to \textsc{reweight}, \citet{Chopin2012a} propose a conditional particle filter (CPF) step to replace $\tilde{\vx}^{1:N_x}_d$ with $\tilde{\vx}^{1:N_x^*}_d$. Here, the state particles and the likelihood estimates are updated by running a particle filter conditional on a single trajectory from the current set of state particles. The incremental weight function of this step is $1$, which means that the parameter particle weights are left unchanged. The drawback of this approach is that all the state particles must be stored, which can significantly increase the RAM required by the algorithm. \citet{Chopin2015} propose two extensions of the CPF approach which reduce the memory requirements of the algorithm at the expense of increased computation time. Their first proposal is to only store the state particles with descendants at the final time-point, i.e.\ using a path storage algorithm within the particle filter \citep{Jacob2015}. Their second method is to store the random seed of the pseudo-random number generator in such a way that the latent states and their associated ancestral indices can be re-generated at any point. Both variants still have a higher RAM requirement and run time compared to the \textsc{reweight} method.
	
	\citet{Duan2014a} propose a reinitialisation scheme to extend the particles (\textsc{reinit}). Whenever $N_x$ is increased, they fit a mixture model $Q(\cdot)$ informed by the current set of particles, then reinitialise the SMC algorithm with $N_x^*$ state particles and $Q(\cdot)$ as the initial distribution. The modified sequence of distributions for \easmcdt{} is 
	\begin{align*}
		\pi_d(\params_d, \vx_{1:T}^{1:N_x} \mid \vy_{1:T}) 
		\propto [Q(\params_d)]^{1-g_d}[p(\params_d)\widehat{p_{N_x}}(\vy_{1:T}\mid\params_d, \vx_{1:T}^{1:N_x})]^{g_d}\psi(\vx_{1:T}^{1:N_x}), \\ \quad 0=g_0\le\cdots\le g_D=1.
	\end{align*}
	The \textsc{reinit} method aims to minimize the variance of the weights, but we find it can be very slow as the algorithm may reinitialise numerous times before completion, each time with a larger number of particles. This approach also assumes that the distribution of the set of parameter particles when \textsc{reinit} is triggered is more informative than the prior, which is not necessarily the case if the adaptation is triggered early. 
	
	\section{Methods} \label{sec:methods}
	This section describes our proposed approach for each of the three stages involved in adapting the number of state particles. 
	
	\subsection{Triggering the adaptation}
	Instead of using the acceptance rate to measure particle diversity, we use the expected squared jumping distance (ESJD), which accounts for both the acceptance rate (the probability that the particles will move) and the jumping distance (how far they will move). See \citet{Pasarica2010}, \citet{Fearnhead2013}, \citet{Salomone2018} and \citet{Bon2021} for examples of this idea outside the SMC$^2$ context. The ESJD at iteration $d$ is defined as
	\begin{align*}
		\textrm{ESJD}_d =\mathbb{E}\left[\norm{\params_d^* - \params_d}^2\right]
	\end{align*}
	where $\norm{\params_d^* - \params_d}^2$ is the squared Mahalanobis distance between the current value of the parameters ($\params_d$) and the proposed value ($\params_d^*$). The ESJD of the $r$th MCMC iteration of the mutation step at iteration $d$ (steps 5-7 of Algorithm \ref{alg:easmc}) can be estimated as 
	\begin{align*}
		\widehat{\textrm{ESJD}}_{d, r} = \frac{1}{N_\theta}\sum_{n=1}^{N_\theta}{(\params_d^n - \params_d^{n,*})^\top\widehat\Sigma^{-1}(\params_d^n-\params_d^{n,*})\alpha(\params_d^n, \params_d^{n,*})},
	\end{align*}
	where $\widehat\Sigma$ is the covariance matrix of the current parameter particle set, and $\alpha(\params_d^n, \params_d^{n,*})$ is the acceptance probability in \eqref{eqn:pmcmc_alpha}. The total estimated ESJD for iteration $d$ is $\widehat{\esjd}_d = \sum_{r=1}^{R}{\widehat{\esjd}_{d, r}}$.
	
	Algorithm \ref{alg:adaptiveMutationStep} outlines how $N_x$ and $R$ are adapted. To summarise, the adaptation is triggered in iteration $d$ if $\widehat{\esjd}_{d-1}$ is below some target value (stage 1). Once triggered, the number of particles is adapted (stage 2) and the particle set is updated (stage 3). A single MCMC iteration is then run with the new number of particles, and the results from this step are used to determine how many MCMC iterations are required to reach the target ESJD, i.e.\ $R$ is given by dividing the target ESJD by the estimated ESJD of the single MCMC iteration and rounding up. Once the adaptation is complete, the remaining MCMC iterations are completed. This approach gives a general framework which can be implemented with any of the stage 2 and stage 3 methods described in Section \ref{sec:easmc_adapting}, as well as our novel methods in Sections \ref{sec:novel_stage2} and \ref{sec:stage3}. 
	
	\begin{algorithm}[htp]
		\small
		
		\begin{adjustwidth}{\algorithmicindent}{}
			\textbf{Input: } the estimated ESJD from the previous iteration ($\widehat{\esjd}_{d-1}$), the target ESJD for each iteration ($\widehat{\esjd}_{\textrm{target}}$) and the current set of particles $\vpar_d^{1:N_{\theta}}$ \\
			\textbf{Output: } new number of state particles $N_x$, estimated ESJD ($\widehat{\esjd}_{d}$) and mutated set of particles $\vpar_d^{1:N_{\theta}}$
		\end{adjustwidth}
		\vspace{0.5em}
		
		\begin{algorithmic}[1]
			\LineComment{Trigger the adaptation}
			\State adapt = $\widehat{\esjd}_{d-1} < \widehat{\esjd}_{\textrm{target}}$
			\vspace{0.5em}
			
			\If{adapt}
			\LineComment{Adapt $N_x$}
			\State \parbox[t]{\dimexpr\textwidth-\leftmargin-\labelsep-\labelwidth}{Set new $N_x$ and update the particle set using any combination of the stage 2 and stage 3 methods described in Sections \ref{sec:easmc_adapting}, \ref{sec:novel_stage2} and \ref{sec:stage3}\strut}
			\EndIf
			\vspace{0.5em}
			
			\LineComment{Initial mutation step with updated $N_x$ (if applicable)}
			\State PMMH mutation $\vpar_{d, 1}^{1:N_{\theta}}\sim K(\vpar_d^{1:N_{\theta}}, \cdot)$, calculate $\widehat{\esjd}_{d, 1}$
			\vspace{0.5em}
			
			\If{adapt}
			\LineCommentIndent{Adapt $R$}	
			\State Set $R = \ceil{\widehat{\esjd}_{\textrm{target}}\slash\widehat{\esjd}_{d, 1}}$
			\EndIf
			\vspace{0.5em}
			
			\LineComment{Remaining mutation steps}
			\For{$r=2$ to $R$} 
			\State PMMH mutation $\vpar_{d, r}^{1:N_{\theta}}\sim K(\vpar^{1:N_{\theta}}_{d, r-1}, \cdot)$ 
			\EndFor
			\vspace{0.5em}
			\State Set $\vpar_d^{1:N_{\theta}} = \vpar_{d ,R}^{1:N_{\theta}}$
			
		\end{algorithmic}
		\caption{Novel method to adapt the number of state particles and mutate the parameter particles for SMC$^2$.}
		\label{alg:adaptiveMutationStep}
	\end{algorithm}

	\subsection{Choosing the new number of particles $N_x^*$} \label{sec:novel_stage2}
	
	To set the new number of state particles $N_x^*$, we build on the \textsc{rescale-var} method of \citet{Duan2014a}, which adapts the number of state particles as follows.
	\begin{enumerate}
		\item Calculate $\bar{\params}_d$, the mean of the current set of parameter samples $\params_d^{1:N_{\theta}}$.
		\item Run the particle filter with $N_x$ state particles $k$ times to get $k$ estimates of the log-likelihood evaluated at $\bar{\params}_d$.
		\item Calculate $\widehat{\sigma_{N_x}}^2$, the sample variance of the $k$ log-likelihood estimates.
		\item Set the new number of state particles to $N_{x}^* = \widehat{\sigma_{N_x}}^2 \cdot N_x$.
	\end{enumerate}
	In practice, we find that \textsc{rescale-var} changes $N_x$ too drastically from one iteration to the next for two reasons. First, the sample variance may itself be highly variable, especially when $N_x$ is small. Second, the sample mean of the parameter particles changes throughout the iterations, meaning that the number of state particles needed to reach a variance of $1$ also changes throughout the iterations. The sample mean may also be a poor value at which to estimate the likelihood if the current target is multimodal or if the current set of parameter particles offers a poor Monte Carlo approximation to the current target distribution. The latter may occur if the number of parameter particles $N_{\theta}$ is too low. 
	
	Our first attempt to overcome some of these problems is to scale the number of state particles by the standard deviation instead of the variance, i.e.\ we set $N_{x}^* = \widehat{\sigma_{N_x}} \cdot N_x$ and call this method \textsc{rescale-std}. A variance of $1$ is still the overall target, however, more moderate values of $N_x$ are proposed when $\widehat{\sigma_{N_x}}^2 \ne 1$. At any given iteration, the new target variance is the current standard deviation, i.e.\ $N_x^*$ is chosen such that $\widehat{\sigma_{N_x^*}}^2 = \widehat{\sigma_{N_x}}$. The main drawback of \textsc{rescale-std} is that the variance at the final iteration may be too high, depending on the initial value of $N_x$ and the variability of the sample variance between iterations, i.e.\ it may approach a variance of $1$ too slowly. In our numerical experiments in Section \ref{sec:ex}, however, we find that the final variance of the \textsc{rescale-std} method is generally between $1$ and $1.2^2$, which is fairly conservative. In their numerical experiments, \citet{Doucet2015} found that the optimal $N_x$ generally gives a variance that is between $1.2^2=1.44$ and $1.5^2=2.25$. 
	
	Our second method (which we refer to as \textsc{novel-var}) aims to improve upon \textsc{rescale-var} by estimating the variance at different values of $N_x$. To obtain our set of candidate values, $\vect{N}_{x, 1:M}$, we scale $N_x$ by different fractional powers of $\widehat{\sigma_{N_x}}^2\slash \sigma_{\textrm{target}}^2$, where $\sigma_{\textrm{target}}^2$ is the target variance. Note that the candidate values $\vect{N}_{x, 1:M}$ will be close to $N_x$ if $\widehat{\sigma_{N_x}}^2$ is close to $\sigma_{\textrm{target}}^2$. To avoid unnecessary computation, the current $N_x$ is left unchanged if $\widehat{\sigma_{N_x}}^2$ falls within some range $\sigma_{\textrm{min}}^2 < \sigma_{\textrm{target}}^2 < \sigma_{\textrm{max}}^2$. We also round the candidate number of state particles up to the nearest $10$, which ensures that there is at least a difference of $10$ between each $N_{x, m} \in \vect{N}_{x, 1:M}$. Once $\vect{N}_{x, 1:M}$ has been obtained, the variance is estimated for each $N_{x, m}\in \vect{N}_{x, 1:M}$, and the new number of state particles is set to the $N_{x, m}$ that has the highest variance less than or equal to $\sigma_{\textrm{max}}^2$. In our numerical experiments in Section \ref{sec:ex}, we set
	\begin{align*}
		\vect{N}_{x, 1:3} = \ceil{N_x \cdot \left\{s^{0.5},\ s^{0.75},\ s \right\}^{\intercal}}, \quad s = \frac{\widehat{\sigma_{N_x}}^2}{\sigma_{\textrm{target}}^2},
	\end{align*}
	which gives candidate values ranging from \textsc{rescale-std} ($s^{0.5} \cdot N_x$) to \textsc{rescale-var} ($s^{1} \cdot N_x$). The target, minimum and maximum variances are $\sigma_{\textrm{target}}^2 = G\cdot 1$, $\sigma_{\textrm{min}}^2 = G\cdot 0.95^2$ and $\sigma_{\textrm{max}}^2 = G\cdot 1.05^2$ respectively, where $G = 1$ for \easmcda{} and $G = 1\slash \max{(0.6^2, g_d^2)}$ for \easmcdt{}. These values are fairly conservative and aim to keep the final variance between $0.95^2\approx0.9$ and $1.05^2\approx1.1$. 
	
	The parameter $G$ is used to take advantage of the effect of the tempering parameter on the variance, i.e.\ $\textrm{var}(\log{(\widehat{p_{N_x}}(\vy\mid\params)^{g_d})}) = g^2 \cdot \textrm{var}(\log{(\widehat{p_{N_x}}(\vy\mid\params))})$. Capping the value of $G$ is necessary in practice, since aiming for an excessive variance is difficult due to the variability of the variance estimate when $N_x$ is low. By setting $G = 1\slash \max{(0.6^2, g_d^2)}$, the highest variance targeted is $1\slash 0.36\approx 2.8$. In general, we recommend not aiming for a variance that is greater than $3$ \citep{Sherlock2015}. Note that including the tempering parameter in this way is infeasible for \textsc{rescale-var} or \textsc{rescale-std}. For the former, changing the target variance only exacerbates the problem of too drastic changes of $N_x$ between iterations. This is largely due to the increased variability of the sample variance when $g_d < 1$. While the variability of $\widehat{\sigma_{N_x}}^2$ is less of a problem for \textsc{rescale-std}, this method struggles keeping up with the increasing variance target. 
	
	Compared to \textsc{rescale-var}, we find that both \textsc{rescale-std} and \textsc{novel-var} are significantly less sensitive to the initial number of state particles, sudden changes in the variance arising from changes in the sample mean of the parameter particles, and variability in the estimated variance of the log-likelihood estimator. The \textsc{novel-var} method is also more predictable in what variance is targeted at each iteration compared to \textsc{rescale-std}.
	
	Our final method (\textsc{novel-esjd}) also compares different values of $N_x$, but using the ESJD instead of the variance of the log-likelihood estimator. As before, the choice of candidate values $\vect{N}_{x, 1:M}$ is flexible, and in the numerical experiments in Section \ref{sec:ex}, we set
	\begin{align}
		\vect{N}_{x, 1:4} = \ceil{N_x \cdot \left\{1, 2, \ s^{0.5},\ s^{1} \right\}^{\intercal}}, \quad s = \frac{\widehat{\sigma_{N_x}}^2}{G},	\label{eq:candidateNx}
	\end{align}
	where $G = 1$ for \easmcda{} and $G = 1\slash \max{(0.6^2, g_d^2)}$ for \easmcdt{}. Again, each $N_{x, m} \in \vect{N}_{x, 1:M}$ is rounded up to the nearest $10$. A score is calculated for a particular $N_{x, m}\in \vect{N}_{x, 1:M}$ by first doing a mutation step with $N_{x, m}$ state particles, then calculating the number of MCMC iterations ($R_m$) needed to reach the ESJD target; the score for $N_{x, m}$ is $(N_m\cdot R_m)^{-1}$. Algorithm \ref{alg:adaptiveMutationStepESJD} describes the adaptive mutation step when using \textsc{novel-esjd}. Since the candidate $N_x$ values are tested in ascending order (see step 2 of Algorithm \ref{alg:adaptiveMutationStepESJD}), it is unnecessary to continue testing the values once the score starts to decrease (steps 8-17 of Algorithm \ref{alg:adaptiveMutationStepESJD}).
	
	This method does not target a particular variance, but instead aims to select the $N_x$ having the cheapest mutation while still achieving the ESJD target. Compared to \textsc{double} and the variance-based methods, we find that \textsc{novel-esjd} is consistent between independent runs, in terms of the run time and the adaptation for $N_x$. It is also relatively insensitive to the initial number of state particles, as well as variability in the variance of the likelihood estimator.
	
	Ideally, the adaptation algorithm (Algorithm \ref{alg:adaptiveMutationStep} or Algorithm \ref{alg:adaptiveMutationStepESJD}) will only be triggered if $N_x$ or $R$ are too low (or too high, as mentioned in Section \ref{sec:ex}). In practice, the ESJD is variable, so the adaptation may be triggered more often than necessary. Allowing the number of state particles to decrease helps to keep the value of $N_x$ reasonable. Also, if the estimated variance is close to the target variance, one of the candidate $N_x$ values will be close in value to the current $N_x$. See Table \ref{tab:candidateNx} for an example of the possible values of $N_x$ for the different methods.
	
	\begin{table}[htp]
		\centering
		\scriptsize
		\begin{tabular}{|c||ccccc|}
			\hline
			$\widehat{\sigma_{N_x}}^2$ & \multicolumn{5}{c|}{Candidate values $\vect{N}_{x}$}  \\
			& \textsc{double} & \textsc{rescale-var} & \textsc{rescale-std} & \textsc{novel-var} & \textsc{novel-esjd} \\
			\hline
			\hline
			0.5 & 200 & 50 & 71 & 50, 60, 71 & 50, 71, 100, 200 \\
			1 & 200 & 100 & 100 & 100 & 100, 200 \\
			1.5 & 200 & 150 & 123 & 123, 136, 150 & 100, 123, 150, 200 \\
			50 & 200 & 5000 & 708 & 708, 1881, 5000 & 100, 200, 708, 5000 \\
			\hline
		\end{tabular}
		\caption{Possible values of the number of state particles $N_x$ if $N_x$ is currently $100$ and $G=1$, where $G$ accounts for the tempering parameter in \easmcdt{}. Note that we allow the number of particles to decrease with \textsc{rescale-var}. The new $N_x$ will be one of the possible values listed, e.g.\ if $\widehat{\sigma_{N_x}}^2 = 1$, \textsc{novel-esjd} will set $N_x$ to $100$ or $200$ depending on which value is predicted to give the cheapest mutation. If there is only $1$ possible value, then that is the new number of state particles.} 
		\label{tab:candidateNx}
	\end{table}
	
	\begin{algorithm}[htp]
		\small
		\begin{adjustwidth}{\algorithmicindent}{}
			\textbf{Input: } the estimated ESJD from the previous iteration ($\widehat{\esjd}_{d-1}$), the target ESJD for each iteration ($\widehat{\esjd}_{\textrm{target}}$) and the current set of particles $\vpar_d^{1:N_{\theta}}$ \\
			\textbf{Output: } new number of state particles $N_x$, estimated ESJD ($\widehat{\esjd}_{d}$) and mutated set of particles $\vpar_d^{1:N_{\theta}}$
		\end{adjustwidth}
		\vspace{0.5em}
		
		\begin{algorithmic}[1]
			\If{$\widehat{\esjd}_{d-1} < \widehat{\esjd}_{\textrm{target}}$}
			\LineCommentIndent{Adapt $N_x$ and $R$}
			\State \parbox[t]{\dimexpr\textwidth-\leftmargin-\labelsep-\labelwidth}{Calculate the set of candidate values, $\vect{N}_{x, 1:M}$ (e.q.\ using \eqref{eq:candidateNx}), and sort in ascending order, such that $N_{x, 1} < N_{x, 2} < \ldots < N_{x, M}$. Set $m^* = M$.\strut}
			\vspace{0.5em}
			
			\For{$N_{x, m}\in \vect{N}_{x, 1:M}$}
			\State \parbox[t]{\dimexpr\textwidth-\leftmargin-\labelsep-\labelwidth}{Replace the current set of state particles with $\tilde{\vx}_d^{1:N_{x, m}}$ using the method described in Section \ref{sec:stage3}\strut}
			\vspace{0.5em}
			
			\State PMMH mutation $\vpar_{d, m}^{1:N_{\theta}}\sim K(\vpar_d^{1:N_{\theta}}, \cdot)$, calculate $\widehat{\esjd}_{d, m}$
			\vspace{0.5em}
			
			\State Calculate $R_m = \ceil{\widehat{\esjd}_{\textrm{target}}\slash\widehat{\esjd}_{d, m}}$ 
			\vspace{0.5em}
			
			\State Calculate score $z_m = (N_{x, m} \cdot R_m)^{-1}$ 
			\vspace{0.5em}
			
			\LineComment{If more than one value has been tested}
			\If{$m > 1$}
			\LineCommentIndent{If the current score is worse than the previous one}
			\If{$z_m\slash z_{m-1} < 1$}
			\State Set $m^* = m-1$ 
			\vspace{0.5em}
			
			\State \parbox[t]{\dimexpr\textwidth-\leftmargin-\labelsep-\labelwidth}{Replace the current set of state particles with $\tilde{\vx}_d^{1:N_{x, m^*}}$ using the method described in Section \ref{sec:stage3}\strut}
			\vspace{0.5em}
			
			\State \textbf{Break}
			\vspace{0.5em}
			
			\LineComment{If the current score is equal to the previous one}
			\ElsIf{$z_m\slash z_{m-1} = 1$}
			\State Set $m^* = m$ 
			\vspace{0.5em}
			\State \textbf{Break}
			\vspace{0.5em}
			\EndIf
			\EndIf
			\EndFor
			\vspace{0.5em}
			\LineComment{Update $N_x$ and $R$}
			\State Set $N_x = N_{x, m^*}$ and $R = R_{m^*}$ 
			
			\Else
			\LineCommentIndent{Initial mutation step}
			\State PMMH mutation $\vpar_{d, 1}^{1:N_{\theta}}\sim K(\vpar_d^{1:N_{\theta}}, \cdot)$, calculate $\widehat{\esjd}_{d, 1}$
			\EndIf
			
			\vspace{0.5em}
			\LineComment{Remaining mutation steps}
			\For{$r=2$ to $R$}
			\State PMMH mutation $\vpar_{d, r}^{1:N_{\theta}}\sim K(\vpar^{1:N_{\theta}}_{d, r-1}, \cdot)$ 
			\EndFor
			\vspace{0.5em}
			\State Set $\vpar_d^{1:N_{\theta}} = \vpar_{d,R}^{1:N_{\theta}}$
			
		\end{algorithmic}
		\caption{Novel method to adapt the number of state particles and mutate the parameter particles for SMC$^2$ when using \textsc{novel-esjd}.}
		\label{alg:adaptiveMutationStepESJD}
	\end{algorithm}

	\subsection{Replacing the state particle set} \label{sec:stage3}
	Our final contribution (denoted \textsc{replace}) is a variation of the \textsc{reweight} scheme of \citet{Chopin2012a}. Both \textsc{reweight} and \textsc{replace} consist of three steps. First, a particle filter (Algorithm \ref{alg:BPF}) is run with the new number of state particles to obtain $\widehat{p_{N_x^*}}(\vy_{1:d}\mid\params_d, \vx_{1:d}^{1:N_x^*})$ and $\vx_{1:d}^{1:N_x^*}$. Second, the parameter particle weights are reweighted using 
	\begin{align*}
		w_d^n = W_d^n \cdot IW_d^n, \quad W_d^n = \frac{w_d^n}{\sum_{i=1}^{N_{\theta}}{w_d^i}},
	\end{align*}
	where $IW_d^n$ is the incremental weight for parameter particle $n$, $n=1, \ldots, N_{\theta}$ at iteration $d$, and finally, the previous likelihood estimate and set of state particles are discarded. Note that prior to this reweighting step, the parameter particles are evenly weighted as the adaptation of $N_x$ is performed after the resampling step, i.e.\ $W_d^n = 1\slash N_{\theta}$, for $n = 1, \ldots, N_{\theta}$. 
	
	With the \textsc{reweight} method, the incremental weights for \easmcda{} are obtained by replacing $p(\vy_{1:d}\mid\params_d)$ with $\widehat{p_{N_x}}(\vy_{1:d}\mid\params_d,\vx_{1:d}^{1:N_x})$ to approximate the optimal backward kernel. This gives
	\begin{align*}
		IW_d = \frac{\widehat{p_{N_x^*}}(\vy_{1:d}\mid\params_d, \vx_{1:d}^{1:N_x^*})}{\widehat{p_{N_x}}(\vy_{1:d}\mid\params_d, \vx_{1:d}^{1:N_x})}.
	\end{align*}
	See Section \ref{sec:easmc_adapting} for details. For \easmcdt, the incremental weights are
	\begin{align*}
		IW_d =\frac{\widehat{p_{N_x^*}}(\vy_{1:T}\mid\params_d, \vx_{1:T}^{1:N_x^*})^{g_d}}{\widehat{p_{N_x}}(\vy_{1:T}\mid\params_d, \vx_{1:T}^{1:N_x})^{g_d}}.
	\end{align*}
	The \textsc{replace} method uses a different approximation to the optimal backward kernel. For \easmcda, instead of using $p(\vy_{1:d}\mid\params_d) \approx p_{N_x}(\vy_{1:d}\mid\params_d,\vx_{1:d}^{1:N_x})$, we use $p(\vy_{1:d}\mid\params_d) \approx p_{N_x^*}(\vy_{1:d}\mid\params_d,\vx_{1:d}^{1:N_x^*})$, which gives the backward kernel 
	\begin{align*}
		L_d(\vx_{1:d}^{1:N_x^*}, \vx_{1:d}^{1:N_x}) &= \frac{
			\widehat{p_{N_x}}(\vy_{1:d}\mid\params_d, \vx_{1:d}^{1:N_x}) \psi(\vx_{1:d}^{1:N_x})
		}{
			p(\vy_{1:d}\mid\params_d)
		} \nonumber \\ &\approx \frac{
			\widehat{p_{N_x}}(\vy_{1:d}\mid\params_d, \vx_{1:d}^{1:N_x}) \psi(\vx_{1:d}^{1:N_x})
		}{
			\widehat{p_{N_x^*}}(\vy_{1:d}\mid\params_d, \vx_{1:d}^{1:N_x^*})
		}.
	\end{align*}
	Using this backward kernel, the incremental weights are
	\begin{align*}
		IW_d = &\frac{
			\pi_d\left(\params_d, \vx_{1:d}^{1:N_x^*} \mid\vy_{1:d}\right)L_d(\vx_{1:d}^{1:N_x^*}, \vx_{1:d}^{1:N_x})
		}{
			\pi_d\left(\params_d, \vx_{1:d}^{1:N_x} \mid\vy_{1:d}\right)\psi(\vx_{1:d}^{1:N_x^*})
		} \\
		&= \frac{
			\widehat{p_{N_x^*}}(\vy_{1:d}\mid\params_d, \vx_{1:d}^{1:N_x^*})
			L_d(\vx_{1:d}^{1:N_x^*}, \vx_{1:d}^{1:N_x})
		}{
			\widehat{p_{N_x}}(\vy_{1:d}\mid\params_d, \vx_{1:d}^{1:N_x}) \psi(\vx_{1:d}^{1:N_x})
		} = 1.
	\end{align*}
	Similarly for \easmcdt, the approximation $p(\vy_{1:T}\mid\params_d)^{g_d} \approx \widehat{p_{N_x^*}}(\vy_{1:T}\mid\params_d,\vx_{1:T}^{1:N_x^*})^{g_d}$ gives the backward kernel 
	\begin{align*}
		L_d(\vx_{1:T}^{1:N_x^*}, \vx_{1:T}^{1:N_x}) &= \frac{
			\widehat{p_{N_x}}(\vy_{1:T}\mid\params_d, \vx_{1:T}^{1:N_x})^{g_d} \psi(\vx_{1:T}^{1:N_x})
		}{
			p(\vy_{1:T}\mid\params_d)^{g_d}
		} \nonumber \\ &\approx \frac{
			\widehat{p_{N_x}}(\vy_{1:T}\mid\params_d, \vx_{1:T}^{1:N_x})^{g_d} \psi(\vx_{1:T}^{1:N_x})
		}{
			\widehat{p_{N_x^*}}(\vy_{1:T}\mid\params_d, \vx_{1:T}^{1:N_x^*})^{g_d}
		}.
	\end{align*}
	and leads to incremental weights
	\begin{align*}
		IW_d = &\frac{
			\pi_d\left(\params_d, \vx_{1:T}^{1:N_x^*} \mid\vy_{1:T}\right)L_d(\vx_{1:T}^{1:N_x^*}, \vx_{1:T}^{1:N_x})
		}{
			\pi_d\left(\params_d, \vx_{1:T}^{1:N_x} \mid\vy_{1:T}\right)\psi(\vx_{1:T}^{1:N_x^*})
		} \\
		&= \frac{
			\widehat{p_{N_x^*}}(\vy_{1:T}\mid\params_d, \vx_{1:T}^{1:N_x^*})^{g_d}
			L_d(\vx_{1:T}^{1:N_x^*}, \vx_{1:T}^{1:N_x})
		}{
			\widehat{p_{N_x}}(\vy_{1:T}\mid\params_d, \vx_{1:T}^{1:N_x})^{g_d} \psi(\vx_{1:T}^{1:N_x})
		} = 1.
	\end{align*}
	Since the incremental weights reduce to $1$, the \textsc{replace} approach introduces no extra variability in the parameter particle weights. As a result, \textsc{replace} leads to less variability in the mutation step compared to the \textsc{reweight} method of \citet{Chopin2012a}, i.e.\ the parameter particles remain evenly weighted throughout the mutation step. We also find that it is generally faster than the \textsc{reinit} method of \citet{Duan2014a}.

	\subsection{Practical Considerations} \label{sec:practical_considerations}
	The framework introduced in this section has a number of advantages over the existing methods. Most notably, the adaptation of $R$ is automated, the stage 2 options (\textsc{rescale-std}, \textsc{novel-var} and \textsc{rescale-esjd}) are less sensitive to variability in the estimated variance of the log-likelihood estimator, and the parameter particle weights are unchanged by adapting $N_x$. 
	
	Two tuning parameters remain to be specified for this method: the target ESJD ($\esjd_{\textrm{target}}$) and the number of samples to use when estimating the variance of the log-likelihood estimator ($k$). In our numerical experiments in Section \ref{sec:ex}, we use $\esjd_{\textrm{target}} = 6$ and $k=100$, which both give reasonable empirical results. The target ESJD has little effect on the value of $N_x$, due to the structure of the updates described in Section \ref{sec:novel_stage2}, but it directly controls $R$. Likewise, $k$ controls the variability of $\widehat{\sigma_{N_x}}^2$. Recall that $\widehat{\sigma_{N_x}}^2$ is the estimated variance of the log-likelihood estimator with $N_x$ state particles and evaluated at the mean of the current set of parameter particles ($\bar{\params}_d$). Ideally, the value of $k$ should change with $N_x$ and $\bar{\params}_d$; however, it is not obvious how to do this. In general, we find that if $\sigma_{N_x}^2\approx \widehat{\sigma_{N_x}}^2$ is high, then the variance of $\widehat{\sigma_{N_x}}^2$ also tends to be high. 
	
	Determining optimal values of $\esjd_{\textrm{target}}$ and $k$ is beyond the scope of this paper, but a general recommendation is to follow \citet{Salomone2018} and set $\esjd_{\textrm{target}}$ to the weighted average of the Mahalanobis distance between the parameter particles immediately before the resampling step. We also recommend choosing $k$ such that the variance of $\widehat{\sigma_{N_x}}^2$ is low ($<0.1$) when $\widehat{\sigma_{N_x}}^2 \approx 1$, i.e.\ the estimate of $\widehat{\sigma_{N_x}}^2$ should have low variance when it is around the target value. This value of $k$ may be difficult to obtain, but again, we find that $k = 100$ gives reasonable performance across all the examples in Section \ref{sec:ex}. To mitigate the effect of a highly variable $\widehat{\sigma_{N_x}}^2$, it is also helpful to set a lower bound on the value of $N_x$, as well as an upper bound if a sensible one is known. An upper bound is also useful to restrict the amount of computational resources that is used by the algorithm.
	
	\section{Examples} \label{sec:ex}
	\subsection{Implementation}
	The methods are evaluated on a simple Brownian motion model, the one-factor stochastic volatility (SV) model in \citet{Chopin2012a}, and two ecological models: the theta-logistic model \citep{Peters2010,Drovandi2022} and the noisy Ricker model \citep{Fasiolo2016a}.
	
	The code is implemented in MATLAB and code is available at https://github.com/imkebotha/adaptive-exact-approximate-smc. The likelihood estimates are obtained using the bootstrap particle filter (Algorithm \ref{alg:BPF}) with adaptive multinomial resampling, i.e.\ resampling is done whenever the effective sample size (ESS) drops below $N_x\slash 2$. The results for all models, except for the Ricker model, are calculated from $50$ independent runs, each with $N_{\theta} = 1000$ parameter samples. Due to time and computational constraints, the Ricker model results are based on $20$ independent runs, each with $N_{\theta} = 400$ parameter samples.
	
	For \easmcdt{}, the temperatures are set adaptively using the bisection method \citep{Jasra2010} to aim for an ESS of $0.6\cdot N_{\theta}$. Similarly, the resample-move step is run for \easmcda{} if the ESS falls below $0.6\cdot N_{\theta}$. As discussed in Section \ref{sec:practical_considerations}, a target \esjd{} of $6$ is used and the sample variance $\widehat{\sigma_{N_x}}^2$ for \textsc{rescale-var}, \textsc{rescale-std}, \textsc{novel-var}, and \textsc{novel-esjd} is calculated using $k = 100$ log-likelihood estimates. For all methods except \textsc{reinit} and \textsc{double}, we also trigger the adaptation whenever $\widehat{\esjd}_{t-1} > 2\cdot\widehat{\esjd}_{\textrm{target}}$ --- this allows the algorithm to recover if the values of $N_x$ and\slash or $R$ are set too high at any given iteration, which may occur e.g.\ with \easmcda{} if there are outliers in the data. When the \textsc{reinit} method is used, a mixture of three Gaussians is fit to the current sample when reinitialising the algorithm. 
	
	The methods are compared based on the mean squared error (MSE) of the posterior mean averaged over the parameters, where the ground truth is taken as the posterior mean from a PMMH chain of length 1 million. As the gold standard ($\textrm{GS}$), \easmcdt{} and \easmcda{} are also run for each model with a fixed number of particles, while still adapting $R$. For each of these runs, the number of state particles is tuned such that $\widehat{\sigma_{N_x}}^2\approx 1$ for the full dataset, and the extra tuning time is not included in the results. 
	
	We use the MSE and the total number of log-likelihood evaluations (denoted TLL) of a given method as a measure of its accuracy and computational cost respectively. Note that each time the particle filter is run for a particular parameter particle, TLL is incremented by $N_x\times t$, where $t$ is the current number of observations. The MSE multiplied by the TLL of a particular method gives its overall efficiency. Scores for the accuracy, computational cost and overall efficiency of a given method relative to the gold standard are calculated as 
	\begin{align*}
		Z_{\textrm{method}, \textrm{MSE}} &:= \frac{\textrm{MSE}_{\textrm{GS}}}{\textrm{MSE}_{\textrm{method}}}, \quad Z_{\textrm{method}, \textrm{TLL}} := \frac{\textrm{TLL}_{\textrm{GS}}}{\textrm{TLL}_{\textrm{method}}}, \\
		Z_{\textrm{method}} &:= Z_{\textrm{method}, \textrm{MSE}}\times Z_{\textrm{method}, \textrm{TLL}}.
	\end{align*}
	Higher values are better. 

	The adaptive mutation step in Algorithm \ref{alg:adaptiveMutationStep} is used for all methods except \textsc{novel-esjd}, which uses the adaptive mutation step in Algorithm \ref{alg:adaptiveMutationStepESJD}. The options for stage 2 are \textsc{double}, \textsc{rescale-var}, \textsc{rescale-std}, \textsc{novel-var} and \textsc{novel-esjd}. Likewise, the options for stage 3 are \textsc{reweight}, \textsc{reinit}, and our novel method \textsc{replace}. Since the aim of the \textsc{novel-var} method is to regularly increase the number of state particles throughout the iterations, the combination \textsc{novel-var} with \textsc{reinit} is not tested. Similarly, due to the number of times $N_x$ is updated when using \textsc{novel-esjd}, only the combination \textsc{novel-esjd} with \textsc{replace} is tested. For all combinations (excluding \textsc{double} and \textsc{reinit}), we allow the number of state particles to decrease. Due to computational constraints, we also cap the number of state particles at $5$ times the number of state particles used for the the gold standard method. Note that the \textsc{double} method cannot decrease $N_x$, and \textsc{reinit} assumes increasing $N_x$ throughout the iterations as the entire algorithm is reinitialised whenever $N_x$ is updated.
		
	To compare the different stage 2 methods, we also plot the evolution of $N_x$ for each example. Recall that $N_x = \mathcal{O}(t)$ for \easmcda{} and $\textrm{var}(\log{(\widehat{p_{N_x}}(\vy\mid\params)^{g_d})}) = g^2 \cdot \textrm{var}(\log{(\widehat{p_{N_x}}(\vy\mid\params))})$ for \easmcdt{}. Based on these two results, a roughly linear increase in $N_x$ is desired --- linear in time for \easmcda{} and linear in $g^2$ for \easmcdt{}. Section \ref{app:post_plots} of the Appendix shows marginal posterior density plots. Section \ref{app:sv_extra} in the Appendix has extra results for the stochastic volatility model with $N_{\theta} = 100$ and $N_{\theta} = 500$, to test the methods with fewer parameter particles.
	
	\subsection{Brownian Motion Model} \label{sec:lgbm}
	The first example is a stochastic differential equation with constant drift and diffusion coefficients,
	\begin{align*}
		dX_t = \left(\beta - \frac{\gamma^2}{2}\right)dt + \gamma dB_t,
	\end{align*}
	where $B_t$ is a standard Brownian motion process \citep[][p.\ 44]{Oksendal2003}. The observation and transition densities are 
	\begin{align*}		
		g(y_t\mid x_{t}, \params) &= \mathcal{N}(x_{t}, \sigma^2), \\
		f(x_t\mid x_{t-1}, \params) &= \mathcal{N}\left(x_{t-1} + \beta - \frac{\gamma^2}{2}, \gamma^2\right). 
	\end{align*}
	One hundred observations are generated from this model using $\params := (x_0, \beta, \gamma, \sigma) = (1, 1.2, 1.5, 1)$ and the priors assigned are $\mathcal{N}(x_0\mid 3, 5^2)$, $\mathcal{N}(\beta\mid 2, 5^2)$, $\operatorname{Half-Normal}(\gamma\mid 2^2)$, and $\operatorname{Half-Normal}(\sigma\mid 2^2)$, respectively. 
	
	Results for all stage 2 and stage 3 combinations are obtained for initial $N_x$ values of $10$ and $100$. The variance of the log-likelihood estimator is around $95$ for $N_x = 10$ and around $2.7$ for $N_x = 100$. The gold standard method is run with $240$ state particles.  
	
	Table \ref{tab:bm_stage3} shows the scores averaged over the two initial values of $N_x$ for the three stage 3 options (\textsc{reweight}, \textsc{reinit} and \textsc{replace}). Note that these scores are relative to \textsc{reweight} instead of the gold standard. Apart from \easmcdt{} with \textsc{double} --- where \textsc{reinit} is faster than \textsc{replace} --- \textsc{replace} consistently outperforms \textsc{reweight} and \textsc{reinit} in terms of statistical and computational efficiency. Interestingly, \textsc{reinit} generally outperforms \textsc{reweight} with \textsc{rescale-std} and \textsc{rescale-var}, but not with \textsc{double}. The performance of \textsc{reinit} greatly depends on the number of times the algorithm is reinitialised and the final number of state particles, and this is generally reflected in the computation time.
	
	Tables \ref{tab:bm_score1} and \ref{tab:bm_score2} show the scores relative to the gold standard for all the \textsc{replace} combinations. \textsc{novel-esjd} has the best overall score followed by \textsc{novel-var} for \easmcdt{}, and \textsc{rescale-var} for \easmcda{}. \textsc{double} performs well on \easmcdt{}, but poorly on \easmcda{} --- it has good statistical efficiency, but is much slower than the other methods. Interestingly, the computational efficiency is generally higher for the adaptive methods than for the gold standard, but their accuracy for \easmcda{} is generally lower. This may be due to high variability in the variance of the log-likelihood estimator and the mean of the parameter particles during the initial iterations of \easmcda{}. Since fewer observations are used to estimate the likelihood in these early iterations ($t < T$), the mean of the parameter particles can change drastically from one iteration to the next, leading to similarly drastic changes in the sample variance of the log-likelihood estimator. 
	
\begin{figure}[htp]
		\centering
		\includegraphics[scale=0.55]{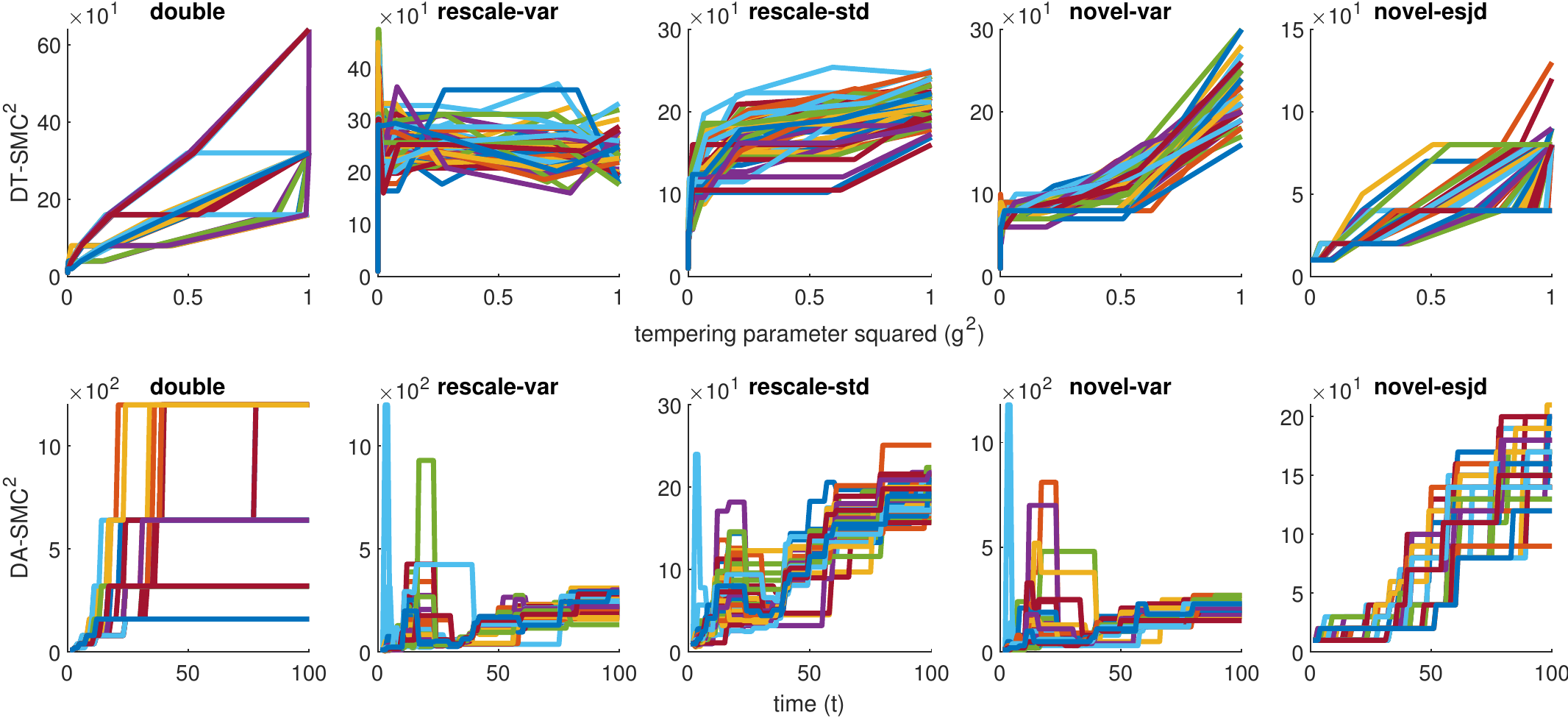}
		\caption{Evolution of $N_x$ for \textsc{replace} and a low initial $N_x$ for the Brownian motion model. Each coloured line represents an independent run of the given method.}
		\label{fig:lgbm_evo}
	\end{figure}

	Figure \ref{fig:lgbm_evo} shows the evolution of $N_x$ for \textsc{replace} and an initial $N_x$ of $10$. Based on these plots, \textsc{double}, \textsc{novel-var} and \textsc{novel-esjd} have the most efficient adaptation for \easmcdt{}, and \textsc{novel-esjd} has the most efficient adaptation for \easmcda{}, which corresponds with the results for $Z_{\textrm{TLL}}$ and $Z$ in Tables \ref{tab:bm_score1} and \ref{tab:bm_score2}.
	
	\begin{table}[htp]
		\centering
		\scriptsize
		\begin{tabular}{|cc||ccc||ccc|}
			\hline
			\multicolumn{2}{|c||}{Method} & \multicolumn{3}{|c||}{\easmcdt{}} & \multicolumn{3}{|c||}{\easmcda{}} \\
			&& $Z_{\textrm{MSE}}$ & $Z_{\textrm{TLL}}$ & $Z$ & $Z_{\textrm{MSE}}$ & $Z_{\textrm{TLL}}$ & $Z$  \\
			\hline
			\hline
			\textsc{double} & \textsc{reweight} & 1.00 & 1.00 & 1.00 & 1.00 & 1.00 & 1.00 \\
			\textsc{double} & \textsc{reinit} & 0.61 & 2.68 & 1.61 & 0.11 & 0.69 & 0.06 \\ 
			\textsc{double} & \textsc{replace} & 1.18 & 1.17 & 1.46 & 1.86 & 1.68 & 2.99 \\ 
			\hline
			\textsc{rescale-var} & \textsc{reweight} & 1.00 & 1.00 & 1.00 & 1.00 & 1.00 & 1.00 \\ 
			\textsc{rescale-var} & \textsc{reinit} & 3.03 & 1.06 & 3.64 & 1.65 & 1.02 & 1.68 \\ 
			\textsc{rescale-var} & \textsc{replace} & 2.97 & 4.76 & 17.46 & 10.59 & 1.91 & 19.49 \\ 
			\hline
			\textsc{rescale-std} & \textsc{reweight} & 1.00 & 1.00 & 1.00 & 1.00 & 1.00 & 1.00 \\ 
			\textsc{rescale-std} & \textsc{reinit} & 5.56 & 1.93 & 11.34 & 1.04 & 1.30 & 1.47 \\ 
			\textsc{rescale-std} & \textsc{replace} & 6.83 & 5.45 & 35.28 & 5.10 & 1.61 & 8.66 \\ 
			\hline
		\end{tabular}
		\caption{Scores for the accuracy ($Z_{\textrm{MSE}}$), computational cost ($Z_{\textrm{TLL}}$) and overall efficiency ($Z$) for the stage 3 options for the Brownian motion model --- higher values are preferred. The results are averaged over the two starting values of $N_x$ and are relative to the \textsc{reweight} method.}
		\label{tab:bm_stage3}
	\end{table}

	\begin{table}[htp]
		\centering
		\scriptsize
		\begin{tabular}{|c||ccc|ccc|}
			\hline
			Method & \multicolumn{6}{c|}{\easmcdt{}} \\
			\hline
			\hline
			Initial $N_x$ &	\multicolumn{3}{c|}{$10$} &  \multicolumn{3}{c|}{$100$} \\
			& $Z_{\textrm{MSE}}$ & $Z_{\textrm{TLL}}$ & $Z$ & $Z_{\textrm{MSE}}$ & $Z_{\textrm{TLL}}$ & $Z$ \\
			\hline
			\hline
			\textsc{gold standard} & 1.00 & 1.00 & 1.00 & 1.00 & 1.00 & 1.00 \\ 
			\textsc{double} & 4.31 & 3.24 & 20.00 & 7.07 & 0.57 & 5.93 \\ 
			\textsc{rescale-var} & 3.32 & 1.21 & 6.24 & 3.68 & 1.21 & 6.71 \\ 
			\textsc{rescale-std} & 4.82 & 2.47 & 18.30 & 4.96 & 1.44 & 10.96 \\ 
			\textsc{novel-var} & 4.21 & 3.26 & 21.01 & 3.89 & 2.43 & 14.41 \\ 
			\textsc{novel-esjd} & 1.95 & 8.75 & 26.34 & 3.58 & 2.42 & 13.16 \\ 
			\hline 
		\end{tabular}
		\caption{Scores for the accuracy ($Z_{\textrm{MSE}}$), computational cost ($Z_{\textrm{TLL}}$) and overall efficiency ($Z$) for \easmcdt{} for the Brownian motion model using the \textsc{replace} method --- higher values are preferred. The gold standard refers to SMC$^2$ with a fixed number of state particles.}
		\label{tab:bm_score1}
	\end{table}

	\begin{table}[htp]
		\centering
		\scriptsize
		\begin{tabular}{|c||ccc|ccc|}
			\hline
			Method & \multicolumn{6}{c|}{\easmcda{}} \\
			\hline
			\hline
			Initial $N_x$ &	\multicolumn{3}{c|}{$10$} &  \multicolumn{3}{c|}{$100$} \\
			& $Z_{\textrm{MSE}}$ & $Z_{\textrm{TLL}}$ & $Z$ & $Z_{\textrm{MSE}}$ & $Z_{\textrm{TLL}}$ & $Z$ \\
			\hline
			\hline
			\textsc{gold standard} & 1.00 & 1.00 & 1.00 & 1.00 & 1.00 & 1.00 \\ 
			\textsc{double} & 1.13 & 0.37 & 0.53 & 1.42 & 0.12 & 0.17 \\ 
			\textsc{rescale-var} & 1.11 & 2.09 & 1.93 & 0.68 & 2.10 & 1.53 \\ 
			\textsc{rescale-std} & 0.50 & 2.52 & 1.36 & 0.58 & 2.27 & 1.44 \\ 
			\textsc{novel-var} & 0.76 & 1.93 & 1.47 & 0.73 & 1.68 & 1.13 \\ 
			\textsc{novel-esjd} & 0.74 & 2.95 & 2.49 & 0.71 & 2.75 & 1.95 \\ 
			\hline 
		\end{tabular}
		\caption{Scores for the accuracy ($Z_{\textrm{MSE}}$), computational cost ($Z_{\textrm{TLL}}$) and overall efficiency ($Z$) for \easmcda{} for the Brownian motion model using the \textsc{replace} method --- higher values are preferred. The gold standard refers to SMC$^2$ with a fixed number of state particles.}
		\label{tab:bm_score2}
	\end{table}

	\subsection{Stochastic Volatility Model}
	Our second example is the one-factor stochastic volatility model used in \citet{Chopin2012a}, 
	\begin{align*}	
		y_t &\sim \mathcal{N}(\mu + \beta v_t, v_t),\\ 
		z_{t} &= \exp{(-\lambda)}z_{t-1} + \sum_{j=1}^k{\exp{(-\lambda(t-c_j))}e_j},\quad z_0\sim\operatorname{Gamma}(\xi^2\slash\omega^2, \xi\slash\omega^2) \\
		v_{t} &=\frac{1}{\lambda}\left[z_{t-1}-z_{t}+\sum_{j=1}^k{e_j}\right], 
		\quad x_{t} = \{v_{t}, z_{t}\}, \\
		k&\sim \operatorname{Poisson}(\lambda\xi^2\slash\omega^2), \quad
		c_{1:k}\overset{\text{iid}}{\sim}\operatorname{Uniform}(t-1, t), \quad e_{1:k}\overset{\text{iid}}{\sim}\operatorname{Exponential}(\xi\slash\omega^2).
	\end{align*}
	The transition density of this model cannot be evaluated point-wise, but it can be simulated from. 
	
	We use a synthetic dataset with $200$ observations, which is generated using $\params := (\xi, \omega^2, \lambda, \beta, \mu)=(4, 4, 0.5, 5, 0)$. The priors are $\operatorname{Exponential}(\xi\mid 0.2)$, $\operatorname{Exponential}(\omega^2\mid 0.2)$, $\operatorname{Exponential}(\lambda\mid 1)$, $\mathcal{N}(\beta\mid 0, 2)$ and $\mathcal{N}(\mu\mid 0, 2)$. 
	
	Results for all stage 2 and stage 3 combinations are obtained for initial $N_x$ values of $300$ and $600$. The variance of the log-likelihood estimator is around $7$ for $300$ state particles and around $3$ for $600$ state particles. The gold standard method is run with $1650$ state particles. 
	
	Table \ref{tab:sv_stage3} shows the scores for the three stage 3 options, relative to \textsc{reweight} and averaged over the two initial $N_x$ values. \textsc{replace} consistently outperforms \textsc{reweight} and \textsc{reinit} in terms of overall efficiency. 
	
	Tables \ref{tab:sv_score1} and \ref{tab:sv_score2} show the scores for all the \textsc{replace} combinations. All methods perform similarly for this model. In terms of accuracy (measured by the MSE), the optimal variance of the log-likelihood estimator seems to be smaller for this model than for the others. However, the efficiency of a smaller variance coupled with the increased computation time is fairly similar to the efficiency of a larger variance with cheaper computation. In this example, \textsc{novel-esjd} has the highest MSE, but the lowest computation time.
	
\begin{figure}[htp]
		\centering
		\includegraphics[scale=0.55]{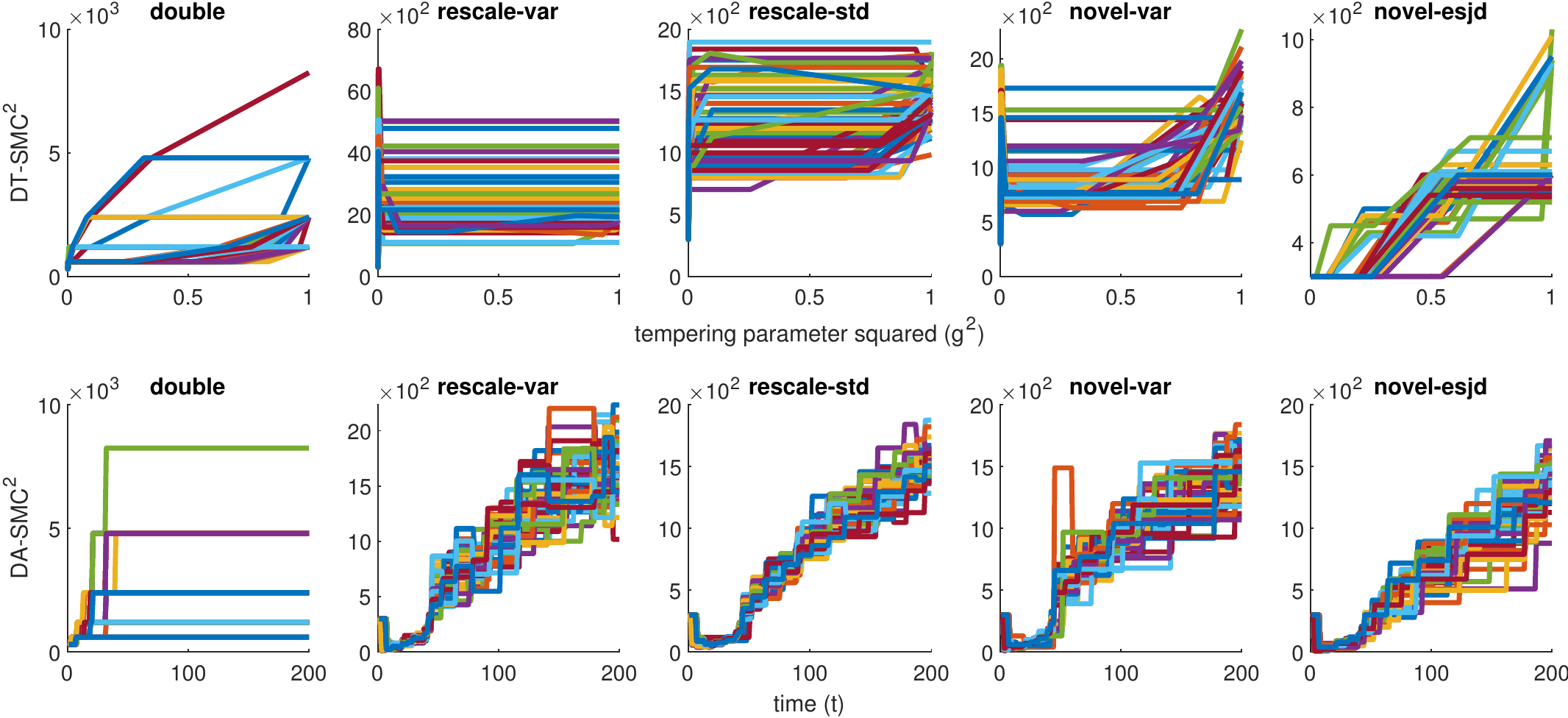}
		\caption{Evolution of $N_x$ for \textsc{replace} and a low initial $N_x$ for the stochastic volatility model. Each coloured line represents an independent run.}
		\label{fig:sv_evo}
	\end{figure}

	Figure \ref{fig:sv_evo} shows the evolution of $N_x$ for \textsc{replace} and an initial $N_x$ of $300$. Based on these plots, \textsc{double} and \textsc{novel-esjd} have the most efficient adaptation for \easmcdt{}, and all methods except \textsc{double} have good results for \easmcda{}. These methods correspond to those with the quickest run time (lowest TLL), but not to the ones with the best overall efficiency.

	\begin{table}[htp]
		\centering
		\scriptsize
		\begin{tabular}{|cc||ccc||ccc|}
			\hline
			\multicolumn{2}{|c||}{Method} & \multicolumn{3}{|c||}{\easmcdt{}} & \multicolumn{3}{|c||}{\easmcda{}} \\
			&& $Z_{\textrm{MSE}}$ & $Z_{\textrm{TLL}}$ & $Z$ & $Z_{\textrm{MSE}}$ & $Z_{\textrm{TLL}}$ & $Z$  \\
			\hline
			\hline
			\textsc{double} & \textsc{reweight} & 1.00 & 1.00 & 1.00 & 1.00 & 1.00 & 1.00 \\
			\textsc{double} & \textsc{reinit} & 1.07 & 0.88 & 0.83 & 0.71 & 0.41 & 0.17 \\ 
			\textsc{double} & \textsc{replace} & 1.38 & 1.15 & 1.48 & 5.09 & 1.10 & 4.31 \\ 
			\hline
			\textsc{rescale-var} & \textsc{reweight} & 1.00 & 1.00 & 1.00 & 1.00 & 1.00 & 1.00 \\ 
			\textsc{rescale-var} & \textsc{reinit} & 4.41 & 1.49 & 7.06 & 0.78 & 0.65 & 0.42 \\ 
			\textsc{rescale-var} & \textsc{replace} & 2.92 & 5.40 & 17.24 & 5.06 & 1.07 & 4.60 \\ 
			\hline
			\textsc{rescale-std} & \textsc{reweight} & 1.00 & 1.00 & 1.00 & 1.00 & 1.00 & 1.00 \\ 
			\textsc{rescale-std} & \textsc{reinit} & 7.33 & 2.07 & 16.21 & 0.26 & 0.44 & 0.13 \\ 
			\textsc{rescale-std} & \textsc{replace} & 4.49 & 5.07 & 24.12 & 1.93 & 1.04 & 1.91 \\ 
			\hline
		\end{tabular}
		\caption{Scores for the accuracy ($Z_{\textrm{MSE}}$), computational cost ($Z_{\textrm{TLL}}$) and overall efficiency ($Z$) for the stage 3 options for the stochastic volatility model --- higher values are preferred. The results are averaged over the two starting values of $N_x$ and are relative to the \textsc{reweight} method}
		\label{tab:sv_stage3}
	\end{table}

	\begin{table}[htp]
		\centering
		\scriptsize
		\begin{tabular}{|c||ccc|ccc|}
			\hline
			Method & \multicolumn{6}{c|}{\easmcdt{}} \\
			\hline
			\hline
			Initial $N_x$&
			\multicolumn{3}{c|}{$300$} &  \multicolumn{3}{c|}{$600$} \\
			& $Z_{\textrm{MSE}}$ & $Z_{\textrm{TLL}}$ & $Z$ & $Z_{\textrm{MSE}}$ & $Z_{\textrm{TLL}}$ & $Z$ \\
			\hline
			\hline
			\textsc{gold standard} & 1.00 & 1.00 & 1.00 & 1.00 & 1.00 & 1.00 \\ 
			\textsc{double} & 1.16 & 2.04 & 2.82 & 1.63 & 1.02 & 1.78 \\ 
			\textsc{rescale-var} & 1.73 & 0.86 & 1.52 & 1.68 & 0.79 & 1.40 \\ 
			\textsc{rescale-std} & 1.26 & 1.66 & 2.20 & 1.35 & 1.27 & 1.75 \\ 
			\textsc{novel-var} & 1.16 & 1.89 & 2.23 & 1.15 & 1.59 & 1.88 \\ 
			\textsc{novel-esjd} & 0.52 & 3.82 & 2.03 & 0.82 & 2.09 & 1.73 \\ 
			\hline 
		\end{tabular}
		\caption{Scores for the accuracy ($Z_{\textrm{MSE}}$), computational cost ($Z_{\textrm{TLL}}$) and overall efficiency ($Z$) for \easmcdt{} for the stochastic volatility model using the \textsc{replace} method --- higher values are preferred. The gold standard refers to SMC$^2$ with a fixed number of state particles.}
		\label{tab:sv_score1}
	\end{table}

	\begin{table}[htp]
		\centering
		\scriptsize
		\begin{tabular}{|c||ccc|ccc|}
			\hline
			Method & \multicolumn{6}{c|}{\easmcda{}} \\
			\hline
			\hline
			Initial $N_x$ &	\multicolumn{3}{c|}{$300$} &  \multicolumn{3}{c|}{$600$} \\
			& $Z_{\textrm{MSE}}$ & $Z_{\textrm{TLL}}$ & $Z$ & $Z_{\textrm{MSE}}$ & $Z_{\textrm{TLL}}$ & $Z$ \\
			\hline
			\hline
			\textsc{gold standard} & 1.00 & 1.00 & 1.00 & 1.00 & 1.00 & 1.00 \\ 
			\textsc{double} & 1.43 & 0.51 & 0.74 & 1.53 & 0.37 & 0.56 \\ 
			\textsc{rescale-var} & 0.80 & 1.34 & 1.06 & 0.71 & 1.33 & 0.96 \\ 
			\textsc{rescale-std} & 0.77 & 1.40 & 1.08 & 0.63 & 1.41 & 0.91 \\ 
			\textsc{novel-var} & 0.75 & 1.38 & 1.05 & 0.91 & 1.38 & 1.28 \\ 
			\textsc{novel-esjd} & 0.63 & 1.35 & 0.89 & 0.67 & 1.31 & 0.88 \\ 
			\hline 
		\end{tabular}
		\caption{Scores for the accuracy ($Z_{\textrm{MSE}}$), computational cost ($Z_{\textrm{TLL}}$) and overall efficiency ($Z$) for \easmcda{} for the stochastic volatility model using the \textsc{replace} method --- higher values are preferred. The gold standard refers to SMC$^2$ with a fixed number of state particles.}
		\label{tab:sv_score2}
	\end{table}

	\subsection{Theta-logistic Model} \label{sec:theta_logistic}
	The next example is the theta-logistic ecological model \citep{Peters2010}, 
	\begin{align*}		
		g(y_t\mid x_{t}, \params) &= \mathcal{N}(y_t\mid a\cdot (x_t), \sigma^2), \\
		x_{t+1} &= x_{t} + \beta_0 + \beta_1\exp{(\beta_2 x_{t})} + z_t, \quad z_t \sim\mathcal{N}(0, \gamma^2).
	\end{align*}
	We fit the model to the first $100$ observations of female nutria populations measured at monthly intervals \citep{Peters2010,Drovandi2022}, using the priors
	$\mathcal{N}(\beta_0\mid 0, 1)$, $\mathcal{N}(\beta_1\mid 0, 1)$, $\mathcal{N}(\beta_2\mid 0, 1)$, $\operatorname{Half-Normal}(\exp{(x_{0})}\mid 1000^2)$, $\operatorname{Exponential}(\gamma\mid 1)$,  $\operatorname{Exponential}(\sigma\mid 1)$ and $\mathcal{N}(a\mid 1, 0.5^2)$.
	
	Scores for the accuracy, computational cost and overall efficiency are obtained for initial $N_x$ values of $700$ and $2400$. The variance of the log-likelihood estimator is around $40$ for $700$ state particles and around $3$ for $2400$ state particles. The gold standard method is run with $4600$ state particles. Due to time constraints, results for the \textsc{double} method with \textsc{reweight} and initial $N_x = 700$ are not available for \easmcda{}.
	
	Table \ref{tab:tl_stage3} shows the scores for the three stage 3 options, averaged over the initial $N_x$ values and relative to \textsc{reweight}. Except for \textsc{double} with \easmcda{}, both \textsc{reinit} and \textsc{replace} outperform \textsc{reweight}, but the results for \textsc{reinit} and \textsc{replace} are mixed. The performance of \textsc{reinit} greatly depends on the number of times the adaptation is triggered. On average, the algorithm is reinitialised fewer times for \textsc{rescale-std} for this example than for the others. 
	
	Tables \ref{tab:tl_score1} and \ref{tab:tl_score2} show the scores for all the \textsc{replace} combinations relative to the gold standard. In this example, \textsc{novel-esjd} outperforms all other methods, followed by \textsc{novel-var} and \textsc{rescale-var}. Unlike the previous examples, \textsc{double} and \textsc{rescale-std} perform poorly here. The gold standard and \textsc{double} have the best MSE for this example, but the worst computation time. The remaining methods have a poor MSE, which is mostly due to the parameter $\sigma$ as Figure \ref{fig:tl_dens} in Section \ref{app:post_plots} of the Appendix shows. The gold standard is the only method that achieves a good result for $\sigma$. 
	
\begin{figure}[htp]
		\centering
		\includegraphics[scale=0.55]{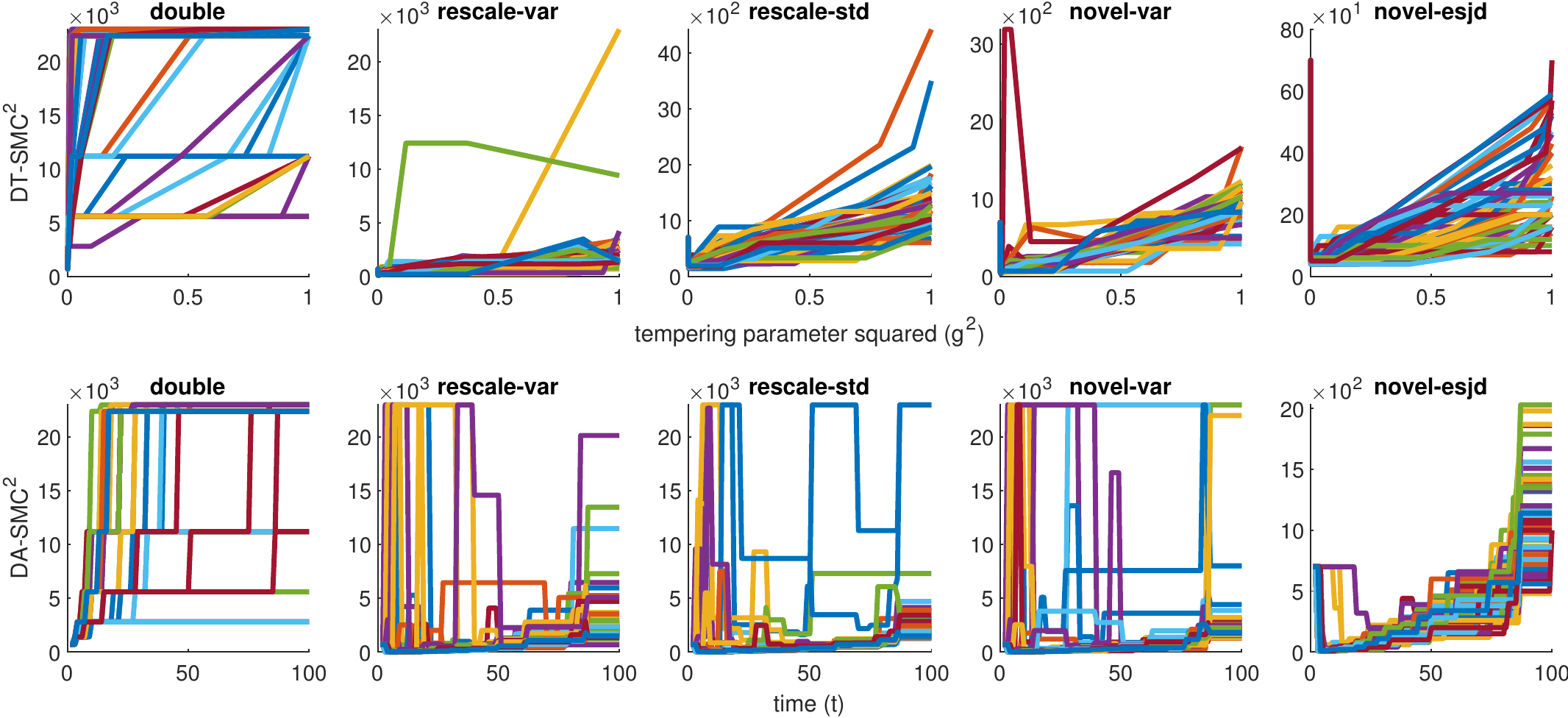}
		\caption{Evolution of $N_x$ for \textsc{replace} and a low initial $N_x$ for the theta-logistic model. Each coloured line represents an independent run.}
		\label{fig:tl_evo}
	\end{figure}

	Figure \ref{fig:tl_evo} shows the evolution of $N_x$ for \textsc{replace} and an initial $N_x$ of $700$. \textsc{novel-esjd} seem to have the least variable evolution for both \easmcdt{} and \easmcda{} compared to the other methods. Again, this is reflected in the values of $Z_{\textrm{TLL}}$, particularly in Tables \ref{tab:tl_score1} and \ref{tab:tl_score2}.

	\begin{table}[htp]
		\centering
		\scriptsize
		\begin{tabular}{|cc||ccc||ccc|}
			\hline
			\multicolumn{2}{|c||}{Method} & \multicolumn{3}{c||}{\easmcdt{}} & \multicolumn{3}{c|}{\easmcda{}} \\
			&& $Z_{\textrm{MSE}}$ & $Z_{\textrm{TLL}}$ & $Z$ & $Z_{\textrm{MSE}}$ & $Z_{\textrm{TLL}}$ & $Z$  \\
			\hline
			\hline
			\textsc{double} & \textsc{reweight} & 1.00 & 1.00 & 1.00 & 1.00 & 1.00 & 1.00 \\
			\textsc{double} & \textsc{reinit} & 0.18 & 7.89 & 1.31 & 0.09 & 1.80 & 0.16 \\ 
			\textsc{double} & \textsc{replace} & 1.18 & 0.94 & 1.11 & 0.85 & 1.09 & 0.89 \\ 
			\hline
			\textsc{rescale-var} & \textsc{reweight} & 1.00 & 1.00 & 1.00 & 1.00 & 1.00 & 1.00 \\ 
			\textsc{rescale-var} & \textsc{reinit} & 0.98 & 6.84 & 7.28 & 0.99 & 1.78 & 1.67 \\ 
			\textsc{rescale-var} & \textsc{replace} & 1.02 & 2.41 & 1.91 & 0.71 & 3.46 & 2.64 \\ 
			\hline
			\textsc{rescale-std} & \textsc{reweight} & 1.00 & 1.00 & 1.00 & 1.00 & 1.00 & 1.00 \\ 
			\textsc{rescale-std} & \textsc{reinit} & 0.99 & 4.14 & 4.24 & 0.76 & 2.42 & 1.78 \\ 
			\textsc{rescale-std} & \textsc{replace} & 1.36 & 1.75 & 3.75 & 0.69 & 3.73 & 2.51 \\ 
			\hline
		\end{tabular}
		\caption{Scores for the accuracy ($Z_{\textrm{MSE}}$), computational cost ($Z_{\textrm{TLL}}$) and overall efficiency ($Z$) for the stage 3 options for the theta-logistic model --- higher values are preferred. The results are averaged over the two starting values of $N_x$ and are relative to the \textsc{reweight} method}
		\label{tab:tl_stage3}
	\end{table}

	\begin{table}[htp]
		\centering
		\scriptsize
		\begin{tabular}{|c||ccc|ccc|}
			\hline
			Method & \multicolumn{6}{c|}{\easmcdt{}} \\
			\hline
			\hline
			Initial $N_x$&
			\multicolumn{3}{c|}{$700$} &  \multicolumn{3}{c|}{$2400$} \\
			& $Z_{\textrm{MSE}}$ & $Z_{\textrm{TLL}}$ & $Z$ & $Z_{\textrm{MSE}}$ & $Z_{\textrm{TLL}}$ & $Z$ \\
			\hline
			\hline
			\textsc{gold standard} & 1.00 & 1.00 & 1.00 & 1.00 & 1.00 & 1.00 \\ 
			\textsc{double} & 1.35 & 0.38 & 0.52 & 1.32 & 0.28 & 0.37 \\ 
			\textsc{rescale-var} & 0.16 & 5.32 & 1.14 & 0.16 & 5.43 & 0.89 \\ 
			\textsc{rescale-std} & 0.13 & 11.23 & 1.49 & 0.14 & 4.39 & 0.76 \\
			\textsc{novel-var} & 0.09 & 20.00 & 1.87 & 0.09 & 9.50 & 1.00 \\ 
			\textsc{novel-esjd} & 0.06 & 34.78 & 2.11 & 0.06 & 19.37 & 1.14 \\ 
			\hline 
		\end{tabular}
		\caption{Scores for the accuracy ($Z_{\textrm{MSE}}$), computational cost ($Z_{\textrm{TLL}}$) and overall efficiency ($Z$) for \easmcdt{} for the theta-logistic model using the \textsc{replace} method --- higher values are preferred. The gold standard refers to SMC$^2$ with a fixed number of state particles.}
		\label{tab:tl_score1}
	\end{table}
	
	\begin{table}[htp]
		\centering
		\scriptsize
		\begin{tabular}{|c||ccc|ccc|}
			\hline
			Method & \multicolumn{6}{c|}{\easmcda{}} \\
			\hline
			\hline
			Initial $N_x$ &	\multicolumn{3}{c|}{$700$} &  \multicolumn{3}{c|}{$2400$} \\
			& $Z_{\textrm{MSE}}$ & $Z_{\textrm{TLL}}$ & $Z$ & $Z_{\textrm{MSE}}$ & $Z_{\textrm{TLL}}$ & $Z$ \\
			\hline
			\hline
			\textsc{gold standard} & 1.00 & 1.00 & 1.00 & 1.00 & 1.00 & 1.00 \\ 
			\textsc{double} & 1.33 & 0.22 & 0.33 & 1.45 & 0.15 & 0.30 \\ 
			\textsc{rescale-var} & 0.25 & 2.17 & 1.09 & 0.24 & 1.86 & 1.00 \\ 
			\textsc{rescale-std} & 0.17 & 2.48 & 0.31 & 0.19 & 2.37 & 0.83 \\
			\textsc{novel-var} & 0.24 & 1.87 & 0.67 & 0.21 & 2.18 & 1.04 \\ 
			\textsc{novel-esjd} & 0.13 & 13.42 & 2.05 & 0.12 & 12.38 & 1.76 \\ 
			\hline 
		\end{tabular}
		\caption{Scores for the accuracy ($Z_{\textrm{MSE}}$), computational cost ($Z_{\textrm{TLL}}$) and overall efficiency ($Z$) for \easmcda{} for the theta-logistic model using the \textsc{replace} method --- higher values are preferred. The gold standard refers to SMC$^2$ with a fixed number of state particles.}
		\label{tab:tl_score2}
	\end{table}

	\subsection{Noisy Ricker Model} \label{sec:ricker}
	Our final example is the noisy Ricker population model \citep{Fasiolo2016a}, 
	\begin{align*}		
		g(y_t\mid x_{t}, \params) &= \operatorname{Poisson}(y_t\mid\phi x_t), \\
		x_{t+1} &= r\cdot x_t \exp{(-x_t + z_{t+1})}, \quad z_t \sim\mathcal{N}(0, \sigma^2).
	\end{align*}
	The transition density of the Ricker model cannot be evaluated point-wise; however, it is straightforward to generate $x_t$ from it, conditional on $x_{t-1}$. This model, and its variants, is typically used to represent highly non-linear or near-chaotic ecological systems, e.g.\ the population dynamics of sheep blowflies \citep{Fasiolo2016a}. \citet{Fasiolo2016a} show that the likelihood function of the noisy Ricker model exhibits extreme multimodality when the process noise is low, making it difficult to estimate the model. 
	
	We draw $700$ observations using $\params := ( \log{(\phi)}, \log{(r)}, \log{(\sigma)}) = (\log{(10)}, \log{(44.7)}, \log{(0.6)})$. Following \citet{Fasiolo2016a}, we assign uniform priors to the log-parameters, $\mathcal{U}(\log{(\phi)}\mid 1.61, 3)$, $\mathcal{U}(\log{(r)} \mid 2, 5)$ and $\mathcal{U}(\log{(\sigma)}\mid -1.8, 1)$, respectively. 
	
	Scores for the accuracy, computational cost and overall efficiency are obtained for initial $N_x$ values of $1000$ and $20000$. The variance of the log-likelihood estimator is around $13$ for $1000$ state particles and around $2.3$ for $20000$ state particles. The gold standard method is run with $90000$ state particles. Due to time constraints, the ground truth for the posterior mean is based on a PMMH chain of length $200000$.
	
	An experiment was stopped if its run time exceeded 9 days. As a result, a full comparison of the stage 3 options cannot be made. Of the experiments that finished, \textsc{replace} had the best results in terms of overall efficiency. On average, \textsc{replace} outperformed \textsc{reinit} and \textsc{reweight} by at least a factor of $2$. In a number of cases, the gold standard and \textsc{replace} were the only methods to finish within the time frame. Tables \ref{tab:ricker_score1} and \ref{tab:ricker_score2} show the scores for the \textsc{replace} combinations. \textsc{novel-var} and \textsc{novel-esjd} have the best overall results across both \easmcdt{} and \easmcda{} for this example, while \textsc{rescale-std} and \textsc{rescale-var} perform similarly.
	
\begin{figure}[htp]
		\centering
		\includegraphics[scale=0.55]{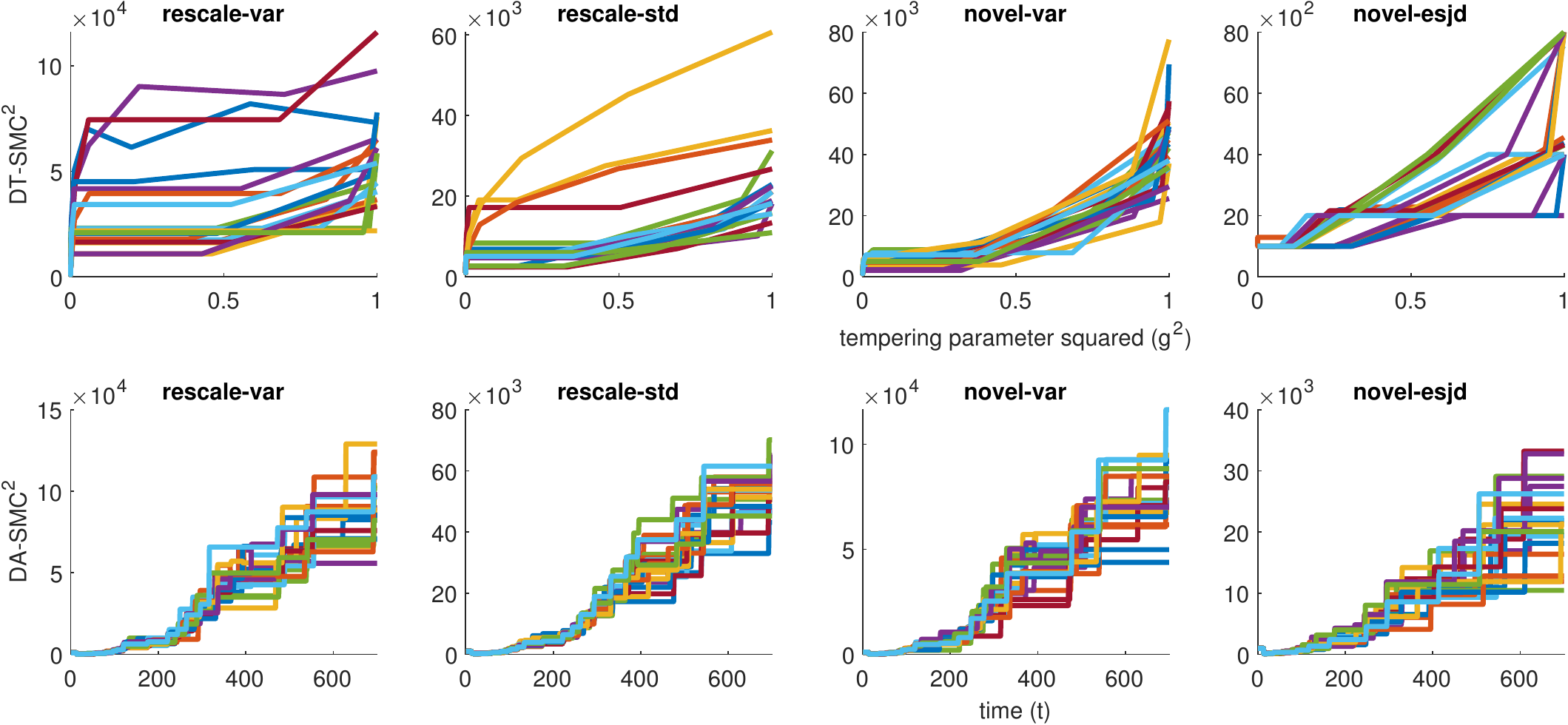}
		\caption{Evolution of $N_x$ for \textsc{replace} and a low initial $N_x$ for the Ricker model. Each coloured line represents an independent run.}
		\label{fig:ricker_evo}
	\end{figure}

	Figure \ref{fig:ricker_evo} shows the evolution of $N_x$ for \textsc{replace} and an initial $N_x$ of $1000$. All methods show a fairly smooth increase in $N_x$ over the iterations.
	
	\begin{table}[htp]
		\centering
		\scriptsize
		\begin{tabular}{|c||ccc|ccc|}
			\hline
			Method & \multicolumn{6}{c|}{\easmcdt{}} \\
			\hline
			\hline
			Initial $N_x$&
			\multicolumn{3}{c|}{$1000$} &  \multicolumn{3}{c|}{$20000$} \\
			& $Z_{\textrm{MSE}}$ & $Z_{\textrm{TLL}}$ & $Z$ & $Z_{\textrm{MSE}}$ & $Z_{\textrm{TLL}}$ & $Z$ \\
			\hline
			\hline
			\textsc{gold standard} & 1.00 & 1.00 & 1.00 & 1.00 & 1.00 & 1.00 \\ 
			\textsc{double} & 0.26 & 12.76 & 3.59 & - & - & - \\ 
			\textsc{rescale-var} & 0.45 & 4.17 & 2.10 & 0.77 & 3.34 & 2.82 \\ 
			\textsc{rescale-std} & 0.33 & 12.79 & 4.62 & 0.54 & 4.28 & 2.40 \\ 
			\textsc{novel-var} & 0.38 & 10.76 & 4.03 & 0.37 & 7.16 & 2.90 \\ 
			\textsc{novel-esjd} & 0.12 & 46.19 & 5.63 & 0.24 & 10.51 & 2.65 \\ 
			\hline 
		\end{tabular}
		\caption{Scores for the accuracy ($Z_{\textrm{MSE}}$), computational cost ($Z_{\textrm{TLL}}$) and overall efficiency ($Z$) for \easmcdt{} for the noisy Ricker model using the \textsc{replace} method --- higher values are preferred. The gold standard refers to SMC$^2$ with a fixed number of state particles.}
		\label{tab:ricker_score1}
	\end{table}
	
	\begin{table}[htp]
		\centering
		\scriptsize
		\begin{tabular}{|c||ccc|ccc|}
			\hline
			Method & \multicolumn{6}{c|}{\easmcda{}} \\
			\hline
			\hline
			Initial $N_x$ &	\multicolumn{3}{c|}{$1000$} &  \multicolumn{3}{c|}{$20000$} \\
			& $Z_{\textrm{MSE}}$ & $Z_{\textrm{TLL}}$ & $Z$ & $Z_{\textrm{MSE}}$ & $Z_{\textrm{TLL}}$ & $Z$ \\
			\hline
			\hline
			\textsc{gold standard} & 1.00 & 1.00 & 1.00 & 1.00 & 1.00 & 1.00 \\ 
			\textsc{double} & - & - & - & - & - & - \\ 
			\textsc{rescale-var} & 0.56 & 2.04 & 1.24 & 0.78 & 2.11 & 1.82 \\ 
			\textsc{rescale-std} & 0.47 & 3.33 & 1.63 & 0.41 & 3.09 & 1.28 \\ 
			\textsc{novel-var} & 0.87 & 2.00 & 1.78 & 1.02 & 2.29 & 2.47 \\ 
			\textsc{novel-esjd} & 0.32 & 6.17 & 2.16 & 0.43 & 5.38 & 2.46 \\ 
			\hline 
		\end{tabular}
		\caption{Scores for the accuracy ($Z_{\textrm{MSE}}$), computational cost ($Z_{\textrm{TLL}}$) and overall efficiency ($Z$) for \easmcda{} for the noisy Ricker model using the \textsc{replace} method --- higher values are preferred. The gold standard refers to SMC$^2$ with a fixed number of state particles.}
		\label{tab:ricker_score2}
	\end{table}

	\section{Discussion} \label{sec:disc}
	We introduce a fully automatic SMC$^2$ algorithm for parameter inference of intractable likelihood state-space models. Of the methods used to select the new number of state particles, \textsc{novel-esjd} gives the most consistent results across all models, choice of initial $N_x$ and between \easmcdt{} and \easmcda{}. This method uses the ESJD to determine which $N_x$ from a set of candidate values will give the cheapest mutation --- this value is selected as the new number of state particles. \textsc{novel-esjd} generally outperforms the other methods in terms of the computational and overall efficiency. A significant advantage of \textsc{novel-esjd} is that the adaptation of $N_x$ is consistent across independent runs of the algorithm (i.e.\ when starting at different random seeds), substantially more so than the other methods. 
	
	Similarly, the \textsc{replace} method typically shows great improvement over \textsc{reweight} and \textsc{reinit}. \textsc{replace} modifies the approximation to the optimal backward kernel used by \textsc{reweight}. This modification means that, unlike \textsc{reweight}, \textsc{replace} leaves the parameter particle weights unchanged. We also find that \textsc{replace} is generally more reliable than \textsc{reinit}. 
	
	Our novel SMC$^2$ algorithm has three tuning parameters that must be set: the target ESJD for the mutation step, the number of log-likelihood evaluations for the variance estimation ($k$) and the initial number of state particles. Determining optimal values of the target ESJD and $k$ is beyond the scope of this paper, but tuning strategies are discussed in Section \ref{sec:practical_considerations}. While any initial number of state particles can be used, a small value yields the most efficient results. Compared to the currently available methods, the new approach requires minimal tuning, gives consistent results and is straightforward to use with both data annealing and density tempering SMC$^2$. We also find that the adaptive methods generally outperform the gold standard, despite the latter being pre-tuned.

	An interesting extension to the current work would be to assess the effect of the target ESJD, the target ESS and the target variance of the log-likelihood estimator when SMC$^2$ is used for model selection. Another area of future work is extending the method for application to mixed effects models \citep{Botha2021}; for these models, it may be possible to obtain significant gains in efficiency by allowing the number of state particles to (adaptively) vary between subjects. The new method can also be used as the proposal function within importance sampling squared \citep{Tran2020}. 
	
	One area of future work is to incorporate more advanced particle filters into our framework, e.g.\ the adaptive particle filters of \citet{Bhadra2016}, \citet{Crisan2018} and \citet{Lee2018}. Another area of future work is to adapt the number of parameter particles ($N_{\theta}$) for a specific purpose, e.g.\ estimation of a particular parameter or subset of parameters. This may reduce the computational resources needed, and applies to SMC methods in general.

	\section{Acknowledgments} \label{sec:ack}
	Imke Botha was supported by an Australian Research Training Program Stipend and a QUT Centre for Data Science Top-Up Scholarship. Christopher Drovandi was supported by an Australian Research Council Discovery Project (DP200102101). We gratefully acknowledge the computational resources provided by QUT's High Performance Computing and Research Support Group (HPC).

	\bibliographystyle{apalike}
	\bibliography{refs}
	
	\appendix
	
	\section{Marginal Posterior Plots} \label{app:post_plots}
	
\begin{figure}[H]  
		\centering
		\includegraphics[scale=0.5]{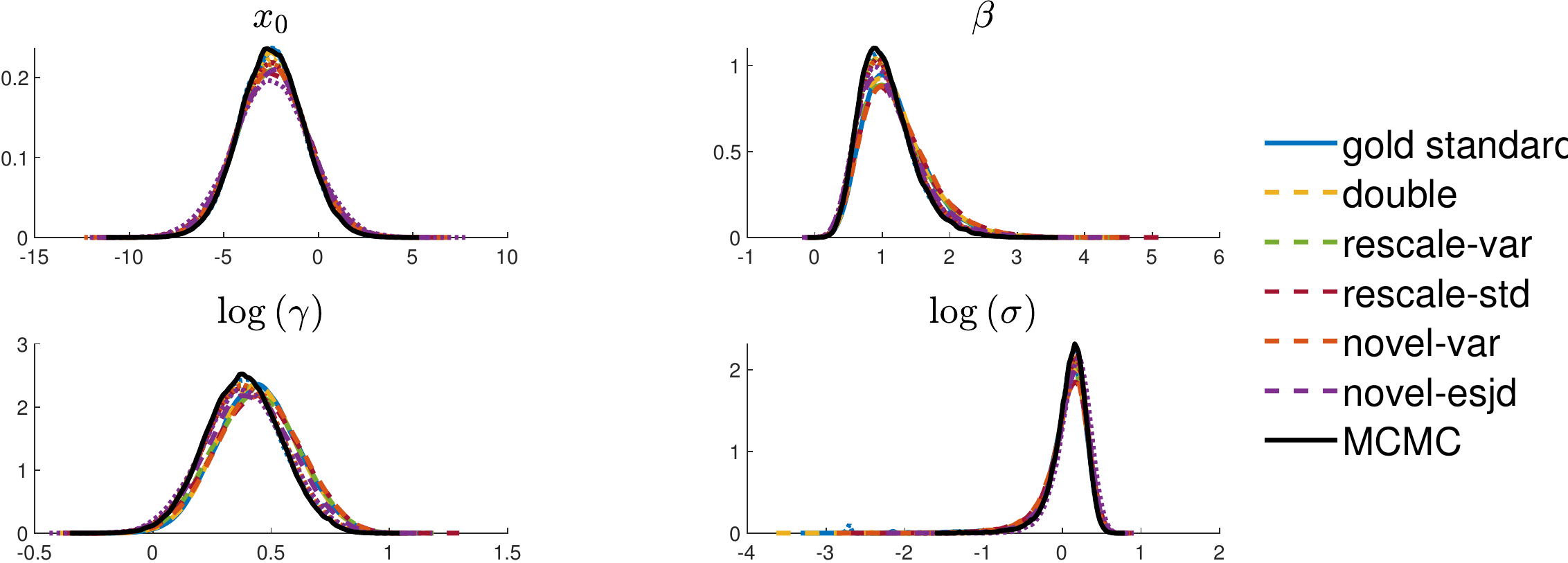}
		\caption{Marginal posterior density plots for the Brownian motion model. Dashed lines are the \easmcda{} results and dotted lines of the same colour are the corresponding \easmcdt{} results.}
		\label{fig:lgbm_dens}
	\end{figure}

\begin{figure}[H]
		\centering
		\includegraphics[scale=0.5]{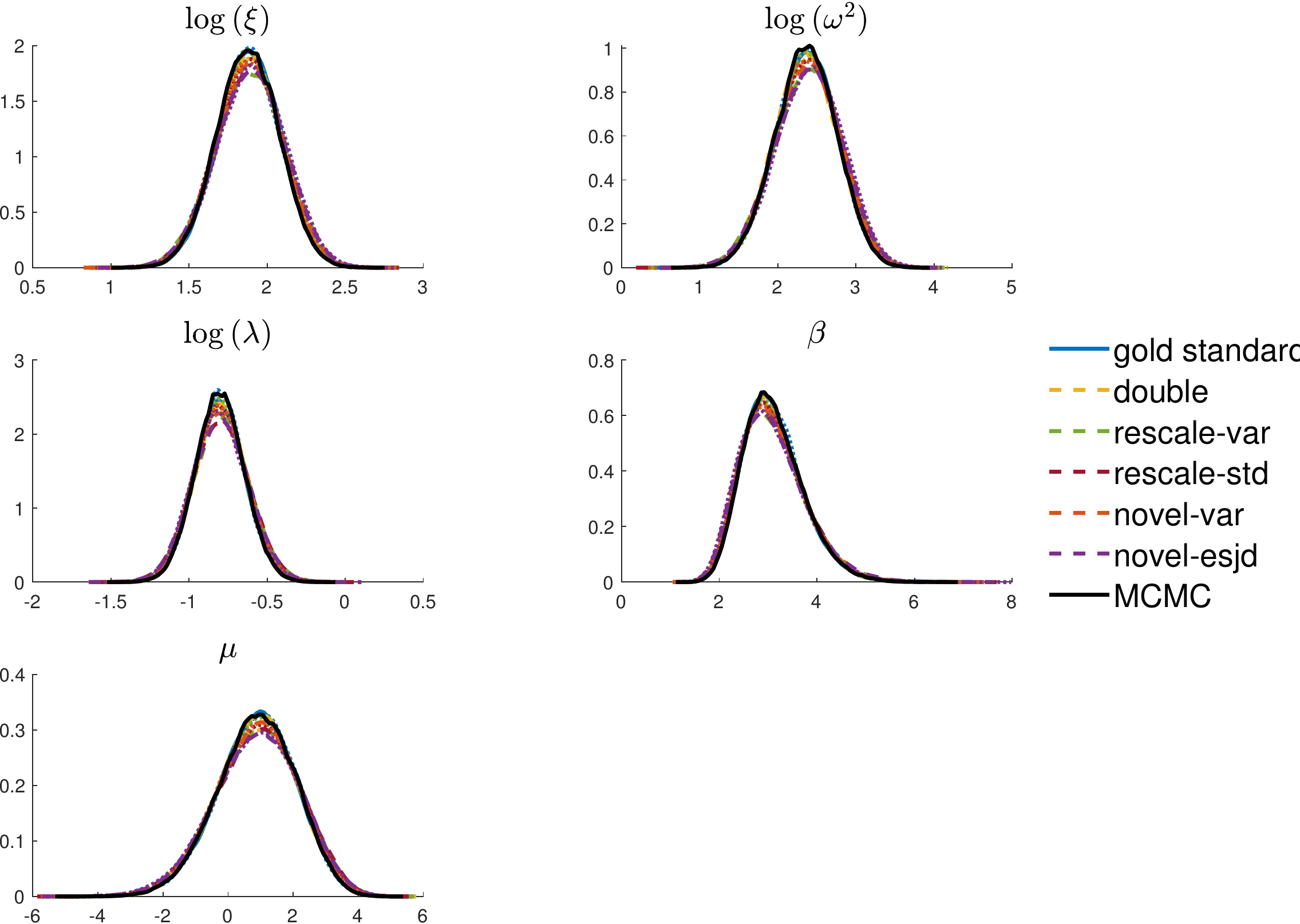}
		\caption{Marginal posterior density plots for the stochastic volatility model. Dashed lines are \easmcda{} results and dotted lines of the same colour are the corresponding \easmcdt{} results.}
		\label{fig:sv_dens}
	\end{figure}

\begin{figure}[H]  
		\centering
		\includegraphics[scale=0.5]{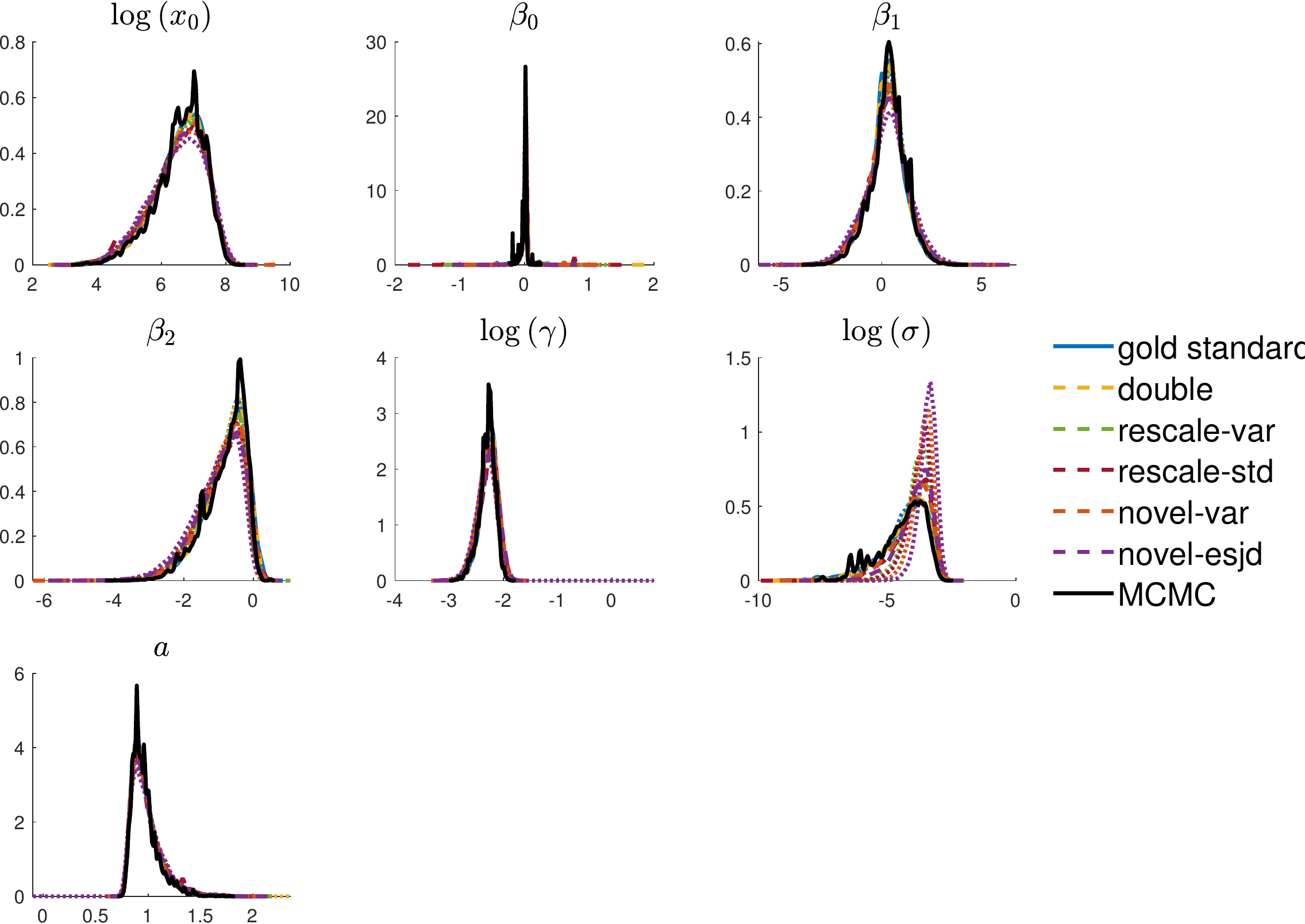}
		\caption{Marginal posterior density plots for the theta-logistic model. Dashed lines are the \easmcda{} results and dotted lines of the same colour are the corresponding \easmcdt{} results.}
		\label{fig:tl_dens}
	\end{figure}

\begin{figure}[H]
		\centering
		\includegraphics[scale=0.5]{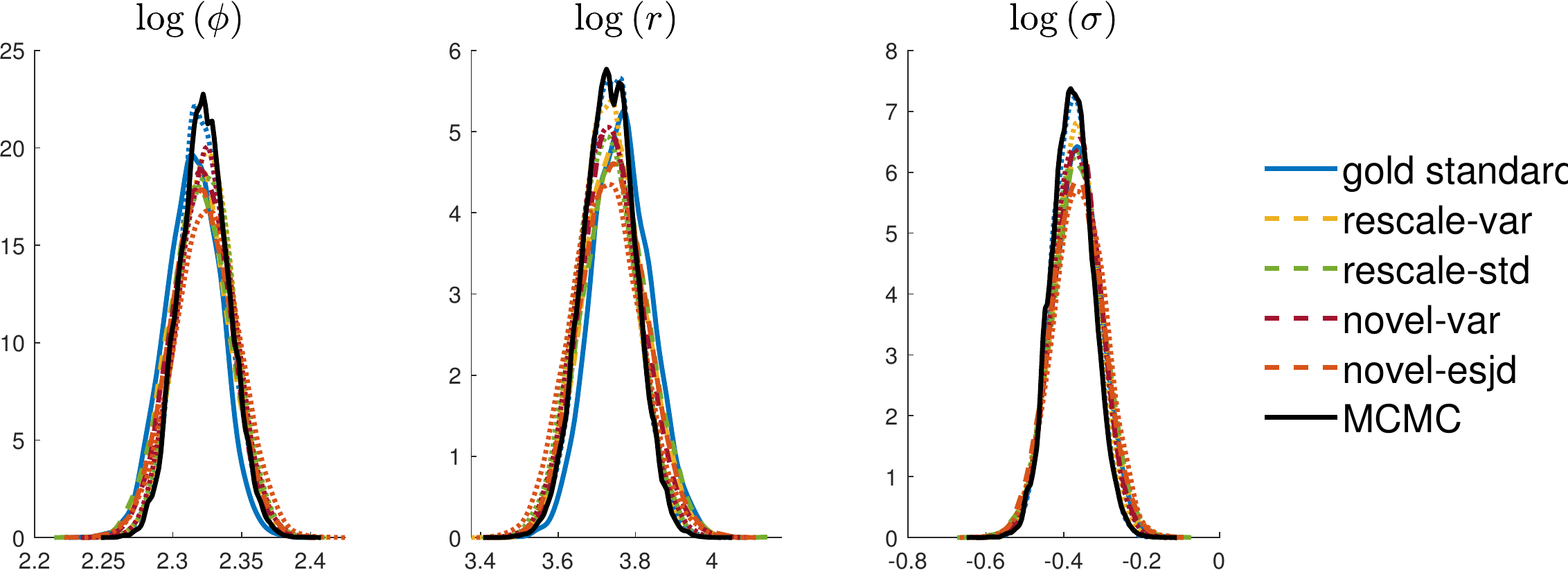}
		\caption{Marginal posterior density plots for the Ricker model. Dashed lines are the \easmcda{} results and dotted lines of the same colour are the corresponding \easmcdt{} results.}
		\label{fig:ricker_dens}
	\end{figure}

	In this section, we show the marginal posterior density plots for the examples in Sections \ref{sec:lgbm}-\ref{sec:ricker}. Figures \ref{fig:lgbm_dens}-\ref{fig:ricker_dens} show the marginal posterior density plots for each example and method. Note that the results shown are for \textsc{replace} using the combined samples from the independent runs, i.e.\ the marginal posteriors are based on $50 \times 1000$ samples for the Brownian motion, stochastic volatility and theta-logistic models and $20 \times 400$ samples for the Ricker model. The results shown are for a low initial $N_x$. It is clear from the plots that the marginal posterior densities are similar between the adaptive methods. The biggest difference in densities are between \easmcdt{} and \easmcda{}, not between the adaptive methods. Figures \ref{fig:lgbm_dens}, \ref{fig:sv_dens} and \ref{fig:ricker_dens} show marginal posteriors from SMC$^2$ that are very similar to the marginal posteriors from MCMC. Figure \ref{fig:tl_dens} shows similar marginal posteriors for the theta-logistic model from SMC$^2$ and MCMC for all of the parameters except for $\log{(\sigma})$. This parameter corresponds to the log of the measurement error in the nutria population data (see Section \ref{sec:theta_logistic} of the main paper). Here, the adaptive SMC$^2$ methods struggle to accurately capture the lower values of $\log{(\sigma})$ with posterior support. SMC$^2$ with a higher, fixed number of state particles (the gold standard method) does not have the same issue, suggesting that the number of state particles is perhaps not adapted high enough in any of the methods for this example.

	\section{Extra Results for the Stochastic Volatility Model} \label{app:sv_extra}
	
	This section shows extra results for the stochastic volatility model. Tables \ref{tab:sv_score1_100} and \ref{tab:sv_score2_100} show the scores for all the \textsc{replace} combinations for $N_{\theta} = 100$, and Tables \ref{tab:sv_score1_500} and \ref{tab:sv_score2_500} show the same results for $N_{\theta} = 500$. There is some variation in the efficiency scores for $N_{\theta} = 100$, $500$ and $1000$, but the results are relatively similar.

	\begin{table}[htp]
		\centering
		\scriptsize
		\begin{tabular}{|c||ccc|ccc|}
			\hline
			Method & \multicolumn{6}{c|}{\easmcdt{}} \\
			\hline
			\hline
			Initial $N_x$&
			\multicolumn{3}{c|}{$300$} &  \multicolumn{3}{c|}{$600$} \\
			& $Z_{\textrm{MSE}}$ & $Z_{\textrm{TLL}}$ & $Z$ & $Z_{\textrm{MSE}}$ & $Z_{\textrm{TLL}}$ & $Z$ \\
			\hline
			\hline
			\textsc{gold standard} & 1.00 & 1.00 & 1.00 & 1.00 & 1.00 & 1.00 \\ 
			\textsc{double} & 0.88 & 1.75 & 1.80 & 1.77 & 0.95 & 1.64 \\ 
			\textsc{rescale-var} & 0.99 & 0.87 & 0.96 & 0.90 & 0.81 & 0.84 \\ 
			\textsc{rescale-std} & 0.96 & 1.62 & 1.68 & 0.75 & 1.22 & 0.92 \\ 
			\textsc{novel-var} & 0.91 & 1.65 & 1.45 & 0.98 & 1.64 & 1.76 \\ 
			\textsc{novel-esjd} & 0.50 & 3.59 & 1.80 & 0.73 & 2.14 & 1.56 \\  
			\hline 
		\end{tabular}
		\caption{Scores for the accuracy ($Z_{\textrm{MSE}}$), computational cost ($Z_{\textrm{TLL}}$) and overall efficiency ($Z$) for \easmcdt{} for the stochastic volatility model with $N_{\theta} = 100$ using the \textsc{replace} method --- higher values are preferred. The gold standard refers to SMC$^2$ with a fixed number of state particles.}
		\label{tab:sv_score1_100}
	\end{table}
	
	\begin{table}[htp]
		\centering
		\scriptsize
		\begin{tabular}{|c||ccc|ccc|}
			\hline
			Method & \multicolumn{6}{c|}{\easmcda{}} \\
			\hline
			\hline
			Initial $N_x$ &	\multicolumn{3}{c|}{$300$} &  \multicolumn{3}{c|}{$600$} \\
			& $Z_{\textrm{MSE}}$ & $Z_{\textrm{TLL}}$ & $Z$ & $Z_{\textrm{MSE}}$ & $Z_{\textrm{TLL}}$ & $Z$ \\
			\hline
			\hline
			\textsc{gold standard} & 1.00 & 1.00 & 1.00 & 1.00 & 1.00 & 1.00 \\ 
			\textsc{double} & 0.90 & 0.56 & 0.62 & 1.55 & 0.36 & 0.54 \\ 
			\textsc{rescale-var} & 0.83 & 1.46 & 1.23 & 0.86 & 1.39 & 1.20 \\ 
			\textsc{rescale-std} & 0.89 & 1.60 & 1.40 & 1.09 & 1.60 & 1.76 \\ 
			\textsc{novel-var} & 1.12 & 1.39 & 1.55 & 1.16 & 1.32 & 1.49 \\ 
			\textsc{novel-esjd} & 0.72 & 1.66 & 1.25 & 0.87 & 1.72 & 1.46 \\ 
			\hline 
		\end{tabular}
		\caption{Scores for the accuracy ($Z_{\textrm{MSE}}$), computational cost ($Z_{\textrm{TLL}}$) and overall efficiency ($Z$) for \easmcda{} for the stochastic volatility model with $N_{\theta} = 100$ using the \textsc{replace} method --- higher values are preferred. The gold standard refers to SMC$^2$ with a fixed number of state particles.}
		\label{tab:sv_score2_100}
	\end{table}
	
	\begin{table}[htp]
		\centering
		\scriptsize
		\begin{tabular}{|c||ccc|ccc|}
			\hline
			Method & \multicolumn{6}{c|}{\easmcdt{}} \\
			\hline
			\hline
			Initial $N_x$&
			\multicolumn{3}{c|}{$300$} &  \multicolumn{3}{c|}{$600$} \\
			& $Z_{\textrm{MSE}}$ & $Z_{\textrm{TLL}}$ & $Z$ & $Z_{\textrm{MSE}}$ & $Z_{\textrm{TLL}}$ & $Z$ \\
			\hline
			\hline
			\textsc{gold standard} & 1.00 & 1.00 & 1.00 & 1.00 & 1.00 & 1.00 \\ 
			\textsc{double} & 1.28 & 1.61 & 2.37 & 1.03 & 1.04 & 1.16 \\ 
			\textsc{rescale-var} & 1.43 & 0.90 & 1.34 & 1.13 & 0.76 & 0.95 \\ 
			\textsc{rescale-std} & 0.80 & 1.81 & 1.52 & 1.02 & 1.20 & 1.25 \\ 
			\textsc{novel-var} & 0.68 & 1.98 & 1.33 & 0.71 & 1.60 & 1.17 \\ 
			\textsc{novel-esjd} & 0.50 & 3.77 & 1.85 & 0.83 & 2.09 & 1.72 \\ 
			\hline 
		\end{tabular}
		\caption{Scores for the accuracy ($Z_{\textrm{MSE}}$), computational cost ($Z_{\textrm{TLL}}$) and overall efficiency ($Z$) for \easmcdt{} for the stochastic volatility model with $N_{\theta} = 500$ using the \textsc{replace} method --- higher values are preferred. The gold standard refers to SMC$^2$ with a fixed number of state particles.}
		\label{tab:sv_score1_500}
	\end{table}
	
	\begin{table}[htp]
		\centering
		\scriptsize
		\begin{tabular}{|c||ccc|ccc|}
			\hline
			Method & \multicolumn{6}{c|}{\easmcda{}} \\
			\hline
			\hline
			Initial $N_x$ &	\multicolumn{3}{c|}{$300$} &  \multicolumn{3}{c|}{$600$} \\
			& $Z_{\textrm{MSE}}$ & $Z_{\textrm{TLL}}$ & $Z$ & $Z_{\textrm{MSE}}$ & $Z_{\textrm{TLL}}$ & $Z$ \\
			\hline
			\hline
			\textsc{gold standard} & 1.00 & 1.00 & 1.00 & 1.00 & 1.00 & 1.00 \\ 
			\textsc{double} & 0.68 & 0.52 & 0.41 & 1.21 & 0.35 & 0.43 \\ 
			\textsc{rescale-var} & 0.86 & 1.39 & 1.22 & 0.71 & 1.39 & 0.98 \\ 
			\textsc{rescale-std} & 0.71 & 1.42 & 1.02 & 0.67 & 1.44 & 1.00 \\ 
			\textsc{novel-var} & 0.75 & 1.39 & 1.03 & 0.92 & 1.38 & 1.29 \\ 
			\textsc{novel-esjd} & 0.92 & 1.43 & 1.35 & 0.66 & 1.39 & 0.93 \\  
			\hline 
		\end{tabular}
		\caption{Scores for the accuracy ($Z_{\textrm{MSE}}$), computational cost ($Z_{\textrm{TLL}}$) and overall efficiency ($Z$) for \easmcda{} for the stochastic volatility model with $N_{\theta} = 500$ using the \textsc{replace} method --- higher values are preferred. The gold standard refers to SMC$^2$ with a fixed number of state particles.}
		\label{tab:sv_score2_500}
	\end{table}

\end{document}